\documentclass[journal]{IEEEtran}
\usepackage{amsmath}
\usepackage{amssymb}
\usepackage{amsfonts}
\usepackage{algorithmic}
\usepackage{bbm}
\usepackage{booktabs}
\newcommand{\Rmnum}[1]{\expandafter\@slowromancap\romannumeral #1@}
\usepackage{graphicx}
\usepackage{epsfig}
\usepackage{subfigure}
\usepackage{citesort}
\usepackage{xcolor}
\usepackage{enumerate}
\usepackage{extarrows}
\makeatletter
\def\ScaleIfNeeded{%
\ifdim\Gin@nat@width>\linewidth \linewidth \else \Gin@nat@width
\fi } \makeatother

\begin{document}

\title{Optimal Throughput Fairness Trade-offs for Downlink Non-Orthogonal Multiple Access over Fading Channels}

\author{Hong Xing, Yuanwei Liu, A. Nallanathan, Zhiguo Ding and H. Vincent Poor
%\thanks{Manuscript received August 07, 2017; revised December 01, 2017; accepted January 20, 2018. This work was supported by the Engineering and Physical Sciences Research Council (EPSRC) of UK under Grant EP/N005651/2. Part of this paper was accepted by IEEE Wireless Communications and Networking Conference (WCNC), Barcelona, Spain, April, 2018 \cite{xing2017WCNC}. The associated editor coordinating the review of this paper and approving it for publication was Prof. Derrick Wing Kwan Ng.}
\thanks{Part of this paper was accepted by IEEE Wireless Communications and Networking Conference (WCNC), Barcelona, Spain, April, 2018 \cite{xing2017WCNC}.}
\thanks{H. Xing, Y. Liu, and A. Nallanathan are with the School of Electronic Engineering and Computer Science, Queen Mary University of London, London, E1 4NS, U.K. (e-mails: h.xing@qmul.ac.uk, \rm yuanwei.liu@qmul.ac.uk, nallanathan@ieee.org).}
\thanks{Z. Ding and H. V. Poor are with the Department of Electrical Engineering, Princeton University, Princeton, NJ 08544, USA (e-mail: poor@princeton.edu).}
\thanks{Z. Ding is also with the School of Computing and Communications, Lancaster University, Lancaster, LA1 4YW, U.K. (e-mail: z.ding@lancaster.ac.uk).}}

\maketitle
\vspace{-.2in}
\begin{abstract}
Recently, {\em non-orthogonal multiple access (NOMA)} has attracted considerable interest as one of the 5G-enabling techniques. However, users with better channel conditions in downlink communications intrinsically benefits from NOMA thanks to successive decoding, judicious designs are required to guarantee user fairness. In this paper, a two-user downlink NOMA system over fading channels is considered. For delay-tolerant transmission, the average sum-rate is maximized subject to both average and peak power constraints as well as a minimum average user rate constraint. The optimal resource allocation is obtained using Lagrangian dual decomposition under full channel state information at the transmitter (CSIT), while an effective power allocation policy under partial CSIT is also developed based on analytical results. In parallel, for delay-limited transmission, the sum of delay-limited throughput (DLT) is maximized subject to a maximum allowable user outage constraint under full CSIT, and the analysis for the sum of DLT is also performed under partial CSIT. Furthermore, an optimal orthogonal multiple access (OMA) scheme is also studied as a benchmark to prove the superiority of NOMA over OMA under full CSIT. Finally, the theoretical analysis is verified by simulations via different trade-offs for the average sum-rate (sum-DLT) versus the minimum (maximum) average user rate (outage) requirement.

\begin{IEEEkeywords}
{N}on-orthogonal multiple access, orthogonal multiple access, fairness, fading channel, ergodic rate, outage probability, Lagrangian dual decomposition, strong duality, 
\end{IEEEkeywords}
\end{abstract}

\IEEEpeerreviewmaketitle
\newtheorem{definition}{\underline{Definition}}[section]
\newtheorem{fact}{Fact}
\newtheorem{assumption}{Assumption}
\newtheorem{theorem}{\underline{Theorem}}[section]
\newtheorem{lemma}{\underline{Lemma}}[section]
\newtheorem{proposition}{\underline{Proposition}}[section]
\newtheorem{corollary}[proposition]{\underline{Corollary}}
\newtheorem{example}{\underline{Example}}[section]
\newtheorem{remark}{\underline{Remark}}[section]
\newcommand{\mv}[1]{\mbox{\boldmath{$ #1 $}}}
\newcommand{\mb}[1]{\mathbb{#1}}
\newcommand{\Myfrac}[2]{\ensuremath{#1\mathord{\left/\right.\kern-\nulldelimiterspace}#2}}

\section{Introduction}
As the incoming fifth generation (5G) wireless communications features massive connectivity among heterogeneous types of users in the Internet of Things (IoT), non-orthogonal multiple access (NOMA) has been envisioned as a promising candidate for 5G networks \cite{Saito13VTC,Fei16MaMIMO,Dai15ComMag}, due to its advantage in enabling high spectral efficiency via non-orthogonal resource allocations over other orthogonal multiple access (OMA) techniques, such as time-division multiple access (TDMA) and frequency-division multiple access (FDMA) (see \cite{Zhiguo2017Mag} and the references therein). Hence, it has recently sparked widespread interest in both industry~\cite{LTE2015,Zhang16TV} and academia~\cite{ding2014performance,Jinho2014Comp,Shin2017NOMA,ding2015cooperative,yuanwei_JSAC_2015,Timotheou:2015,Yuanwei2016NOMA,chen17mathematical,Di2016TWC,Sun17FD-NOMA}. Variation forms of NOMA, namely, multi-user superposition transmission (MUST) and layer division multiplexing (LDM), have been included in the 3rd Generation Partnership Project Long Term Evolution Advanced (3GPP-LTE-A)~\cite{LTE2015} and the next general digital TV standard ATSC 3.0~\cite{Zhang16TV}, respectively.

Among a variety of studies addressing the challenges posed by NOMA, a general NOMA downlink framework was proposed in~\cite{ding2014performance} in which a base station (BS) is capable of simultaneously communicating with several randomly deployed users. 
%In~\cite{Ali2016access}, the user clustering and power allocation problems were studied for both downlink and uplink NOMA transmissions demonstrating the differences in the working principles. 
To increase the throughput of cell-edge users in multi-cell NOMA networks, coordinated multi-point (CoMP) transmission techniques were adopted in~\cite{Jinho2014Comp} and~\cite{Shin2017NOMA} with the BS equipped with a single antenna and multiple antennas, respectively.

On another front, far users that suffer from severe path-loss attenuation are usually disadvantaged in competing for resources enhancing the sum throughput of the system, and therefore their performance could be substantially compromised without proper design. Multi-user proportional fairness was adopted in \cite{Saito13VTC} as a scheduling metric to achieve a good trade-off between system throughput and user fairness. There are mainly three types of countermeasures against such unfairness in NOMA networks. The \emph{first strategy} is to invoke cooperative NOMA~\cite{ding2015cooperative,yuanwei_JSAC_2015}, in which a nearby user is regarded as a relay to assist a distant user. It is demonstrated in~\cite{ding2015cooperative} that by utilizing the proposed cooperative protocol, all users experience the same diversity order. In \cite{yuanwei_JSAC_2015} the nearby NOMA users are equipped with wireless energy harvesting capability to assist far users. 
%In view of possible extra bandwith invoked in cooperative transmission, the full duplex technique was proposed to invoke at the nearby user in \cite{ZhangTVT2016}, for further enhancing the spectral efficiency of cooperative NOMA.
The \emph{second strategy} is to enhance the worst user performance \cite{Timotheou:2015,Yuanwei2016NOMA,chen17mathematical}. The max-min power allocation problem that maximizes the minimum achievable user rate was studied for single-input-single-output (SISO) NOMA systems in~\cite{Timotheou:2015}, and for clustered multiple-input-multiple-output (MIMO) NOMA systems in~\cite{Yuanwei2016NOMA}. In \cite{chen17mathematical}, the authors provided a mathematical proof for NOMA's superiority over conventional OMA transmission in terms of the optimum sum rate subject to a minimum rate constraint. 
The \emph{third strategy} is to introduce additional factors to guarantee fairness. Weighted sum-rate is an effective metric to reflect the priority of users in resource allocation \cite{Di2016TWC,Sun17FD-NOMA}. \cite{Di2016TWC} considered a mutli-carrier downlink network, in which each sub-channel can be shared by multiple users by adopting NOMA. Joint sub-channel and power allocation was formulated as a weighted sum-rate maximization problem, and iteratively solved by leveraging a matching problem with externalities. multi-carrier NOMA systems employing a full-dupex (FD) BS was considered in \cite{Sun17FD-NOMA}, and an optimal joint sub-carrier and power policy for maximizing the weighted sum-rate was developed by applying monotonic optimization. 
\vspace{-.15in}
\subsection{Related Work}
Fairness issues were studied for NOMA over fading channels in the above work. However, they were either considered in a long term with fixed power allocations, e.g., in \cite{ding2014performance,yuanwei_JSAC_2015}, or investigated exploiting adaptive allocation of power and/or bandwidth in a short term, e.g., in \cite{Di2016TWC,Sun17FD-NOMA}. By contrast, we consider adaptive resource allocations to channel dynamics for a two-user downlink NOMA over the whole fading process, the system design of which requires satisfying long-term constraints for quality-of-service (QoS) thus posing new challenges compared with short-term objectives. The information theoretic study of fading broadcast channels (BCs) can be traced back to \cite{Lifang01EC} and \cite{Lifang01OC}. Assuming perfect channel state information (CSI) at both the transmitter (Tx) and the receivers (Rxs), dynamic power and rate allocations for various transmission schemes including code division (CD) with and without successive decoding, time division, and frequency division over different fading states were studied for the ergodic capacity region (ECR) and the (zero-) outage capacity region (OCR) in~\cite{Lifang01EC} and \cite{Lifang01OC}, respectively. The boundaries of the ECRs have been characterized in \cite{Lifang01EC} by solving equivalent weighted sum-rate problems each corresponding to one set of weights. The (zero-) OCRs were inexplicitly characterized by deriving the outage probability regions given a rate vector in \cite{Lifang01OC}. The boundaries of these regions were also obtained by solving equivalent sum-reward maximization problems \cite{Lifang01OC}. 

\textcolor{black}{While \cite{Lifang01EC} studied the boundary of the ergodic capacity region by solving an equivalent average weighted-sum rate problem subject to an average total power constraint, the optimal throughput fairness trade-off region that we characterize in this paper is obtained by maximizing the average sum rate subject to a minimum average rate constraint in addition to average and/or peak power constraints. Therefore, the single-variable Lagrangian multiplier employed to decide the ``water-filing'' power level therein is not readily applicable to our proposed problem. With more Lagrangian multipliers involved, our formulated Lagrangian can be decoupled into many (equal to the total number of fading states) subproblems, which can thus be solved in a parallel fashion  with high efficiency. On the other hand, in \cite{Lifang01OC}, assuming that the transmission to each user is independent, for each joint fading state, an outage was declared when a given rate vector cannot be maintained for all the users using CD either with or without successive decoding. By contrast, we considered a more general scenario in which, for example, under full CSI at the Tx (CSIT), even if the user with better channel condition fails to decode the weak user's message using successive decoding, it is still possible to directly retrieve its own  treating the interference from the weak user as noise. Furthermore, unlike \cite{Lifang01OC} that defined the usage probability via the power set of the users, we equivalently reformulate this continuous variable by arithmetic operation over multiple discrete variables via an indicator function \cite{xing2016secrecySWIPT}.}
\vspace{-.25in}
\subsection{Motivation and Contributions}
\textcolor{black}{Since the performance of users with disadvantage channel conditions over multi-user fading BC tends to be compromised for the objective of mere sum-throughput maximization, we aim for maximizing the sum throughput of these systems while satisfying the QoS of the worst user. The classical results derived in the above work are nevertheless not readily extendible to problems with minimum ergodic rate constraints in delay-tolerant scenarios or those with maximum outage constraints in delay-limited scenarios. Although \cite{Nihar03minimum} investigated the minimum-rate capacity region taking fairness into account, it imposes the minimum rate constraint in every fading state, which may require quite complex encoding/coding design (see Section IV. B of \cite{Nihar03minimum}). Furthermore, other than the modified ``water-filling'' based optimal power allocation procedure \cite{Dirk1975thesis,Lifang01EC} that requires iteratively selecting the ``best'' Rx for each fading state, in this paper we are interested in optimal solution that can be obtained more efficiently, e.g., by solving a series of subproblems in parallel.}

Motivated by these new challenges, we study the average sum-rate and/or the sum of delay-limited throughput (DLT) maximization subject to user fairness for a two-user downlink  NOMA system over fading channels. The main contributions of this paper are summarized as follows. We 1) solve the ergodic sum-rate (ESR) maximization problem ensuring a minimum average user rate by optimally adapting the power and rate allocations to fading states with full CSIT for both NOMA and an optimal OMA scheme; 2) obtain the optimal power control to the sum of DLT maximization problem, which is subject to a maximum permissive user outage constraint, with full CSIT for both NOMA and the optimal OMA scheme; 3) under full CSIT, prove the superiority of NOMA over OMA in terms of the considered metrics; 4) under partial CSIT, analyse the ESR and the DLT, respectively, in closed-form with the static power allocation and/or proportion of orthogonal resources designed; and 5) characterize the optimal average sum-rate (sum-DLT) versus min-rate (max-outage) trade-offs for different transmission schemes via simulations. 

The remainder of the paper is organized as follows: Section~\ref{sec:System Model} introduces the system model and the corresponding performance metrics. In Section~\ref{sec:Optimum Delay-Tolerant Transmission}, the average sum-rate is maximized subject to transmit power constraints as well as a minimum average user rate constraint under full and partial CSIT, respectively, while in Section~\ref{sec:Optimum Delay-Limited Transmission}, the sum of DLT is maximized subject to transmit power constraints as well as a maximum user outage constraint. Numerical results are provided in \ref{sec:Numerical Results}. Finally, Section~\ref{sec:Conclusion} concludes the paper.
	 
{\it Notation}---We use upper-case boldface letters for matrices and lower-case boldface letters for vectors. $\nabla_{\mv x}f(\mv x)$ denotes the gradient of \(f(\mv x)\) with respect to (w.r.t.) $\mv x$. $\mb{E}_x[\cdot]$ stands for the statistical expectation w.r.t. the random variable (RV) $x$. $\sim$ represents ``distributed as'' and $\triangleq$ means ``denoted  by''. The circularly symmetric complex Gaussian (CSCG) distribution with mean \(u\) and variance \(\sigma^2\) is denoted by \(\mathcal{CN}(u,\sigma^2)\). \({\rm Ei}(x)=\int_{-\infty}^{x}\frac{e^t}{t}\mathrm{d}t\) (\(x<0\)) is the exponential integral function of argument \(x\). In addition, \((x)^+=\max(0,x)\) and \([x]_a^b=\max(\min(x,b),a)\).
 
\section{System Model}\label{sec:System Model}
\begin{figure}[htp]
	\begin{center}
		\scalebox{0.25}{\includegraphics*{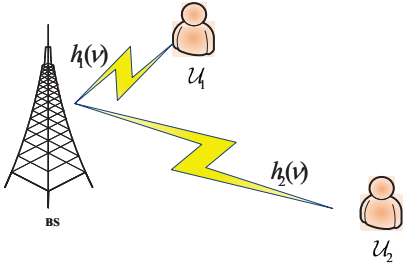}}
	\end{center}
	\caption{System model for a two-user downlink NOMA.} \label{fig:system model}
\end{figure}

We consider a simplified single-carrier downlink cellular system that consists of one BS and two users\textcolor{black}{\footnote{We consider a single-carrier multi-user downlink NOMA with only two users for the following two factors. First, the two-user case is practically favourable to industry \cite{MUST2015}, since the delay incurred in successive interference cancellation (SIC) is significantly reduced. Second, insights for system design can be drawn easily from the two-user solution, while general solution with more than two users can also be obtained under full CSIT without much difficulty. In addition, more complex design for multi-carrier NOMA can be applied to each transmission block considered herein, but is beyond the scope of this paper. The interested reader can refer to \cite{Di2016TWC,Sun17FD-NOMA} for multi-carrier based transmission schemes in NOMA.}}, denoted by \(\mathcal{U}_k\), \(k\in\{1,2\}\), as shown in Fig. \ref{fig:system model}. Both the BS and the users are assumed to be equipped with single antenna. We assume that the complex channel coefficient from the BS to \(\mathcal{U}_k\), \(h_k(\nu)\) experiences block fading
%\textcolor{black}{\footnote{A more complex multi-carrier system model can be considered, e.g., by applying the joint resource allocation algorithms proposed in \cite{Yan2016GC,Zhiqiang17MC-NOMA} to each fading state in ours. This type  of multi-carrier NOMA fading BC is nevertheless beyond the discussion of this paper, and is left for future studies.}}
with a continuous joint probability density function (pdf), where \(\nu\) represents a fading state. The channel remains constant during each transmission block, but may vary from block to block as \(\nu\) changes\textcolor{black}{\footnote{Note that the ``block fading'' herein refers to slow fading scenarios in which the channel remains constant within each block length such that short-length coding schemes are applicable.}}. The channel gain \(|h_k(\nu)|^2\) is assumed to consist of multiplicative small scale and large scale fading given by \(|h_k(\nu)|^2=\frac{|\bar{h}_k(\nu)|^2}{\lambda_k}\), in which \(\bar{h}_k(\nu)\) is a complex Gaussian RV denoted by \(\bar{h}_k(\nu)\sim\mathcal{CN}(0,1)\), and \(\lambda_k\) is a distant-dependent constant. Hence, \(|h_k(\nu)|^2\) is an exponentially distributed RV with its mean value specified by \(\Myfrac{1}{\lambda_k}\).

\subsection{Full CSIT}
In this paper, we investigate two types of CSIT, i.e., full CSIT and partial CSIT, while CSI at the Rxs is assumed to be perfectly known. When full CSIT is available, the BS can adapt its power and rate of the transmit signal intended for each user to the channel \(h_k(\nu)\)'s in each fading state. On the other hand, when only partial CSIT including the order of the two channel gains and their channel distribution information (CDI) is available, due to some reasons like limited feedback from the users to the BS or reducing signalling for the purpose of reducing overhead, the BS can only determine its power allocation policy at each fading state based on this order. We also consider two different multiple access transmission schemes, viz., NOMA and optimal OMA. In the NOMA transmission scheme, the two users non-orthogonally access the channel by enabling superposition coding (SC) at the BS and SIC at the users. For optimal OMA transmission, we consider power and (continuous) time/frequency allocation both in an adaptive manner, which is referred as {\em OMA-TYPE-II} \cite{chen17mathematical}. (Another benchmark scheme, {\em OMA-Type-I}, will be introduced in Section \ref{sec:Numerical Results}.)   

\subsubsection{NOMA}
For NOMA transmission, the received signal at the downlink user \(\mathcal{U}_k\) is given by
\begin{align}
y_k(\nu)=\sqrt{p_k(\nu)}h_k(\nu)s_k+\sqrt{p_{\bar k}(\nu)}h_k(\nu)s_{\bar k}+n_k, 
\end{align}
where \(\bar{k}\) denotes the element in the complementary set of \(\{1,2\}\) w.r.t. \(k\); \(s_k\)'s is the transmit signal intended for \(\mathcal{U}_k\)'s, denoted by \(s_k\sim\mathcal{CN}(0,1)\)\textcolor{black}{\footnote{Note that in real communications system with transmitted signals drawn from finite-alphabet (i.e., discrete) constellations and uniform distribution, the associated encoding/decoding schemes must be judiciously designed such that SIC detector is performed to satisfied level \cite{yongpeng15finite-alphabet,Zheng17JSAC}. However, the associated design is beyond the scope of this paper, and is left as an interesting future direction.}}; \(p_k(\nu)\)'s denotes \(\mathcal{U}_k\)'s transmit power; and \(n_k\)'s is the AWGN at \(\mathcal{U}_k\)'s Rx, denoted by \(n_k\sim\mathcal{CN}(0,\sigma_k^2)\). 

We also define \(\mathcal{U}_k\)'s achievable rate for decoding \(\mathcal{U}_{\bar k}\)'s message at fading state \(\nu\) in bits/sec/Hz, \(k\in\{1,2\}\), treating interference as noise (TIAN), as follows.
\begin{align}
R_{k\rightarrow\bar{k}}^{\rm NOMA}(\nu)=\log_2\left(1+\frac{p_{\bar k}(\nu)g_k(\nu)}{p_k(\nu)g_k(\nu)+1} \right), \label{eq:achievable rate for k to decode bar k}
\end{align} where \(g_k(\nu)\) is the normalized channel gain given by \(\frac{|h_k(\nu)|^2}{\sigma_k^2}\triangleq g_k(\nu)\). 
Similarly, the achievable rate for \(\mathcal{U}_k\) to decode its own message by TIAN is given by
\begin{align}
R_{k\rightarrow k}^{\rm NOMA}(\nu)=\log_2\left(1+\frac{p_k(\nu)g_k(\nu)}{p_{\bar k}(\nu)g_k(\nu)+1} \right). \label{eq:achievable rate for k to decode k}
\end{align}
If \(g_k(\nu)>g_{\bar k}(\nu)\), it implies that \(R_{k\rightarrow\bar{k}}^{\rm NOMA}(\nu)>R_{\bar{k}\rightarrow\bar{k}}^{\rm NOMA}(\nu)\). In other words, under this condition, the achievable rate for \(\mathcal{U}_k\) (the stronger user) to decode the message of \(\mathcal{U}_{\bar k}\) (the weaker user) is larger than that intended for \(\mathcal{U}_{\bar k}\)'s transmission, and therefore \(\mathcal{U}_k\) is able to successfully perform SIC. Otherwise, \(\mathcal{U}_k\) is only able to decode its own message by TIAN. To sum up, the instantaneous achievable rate for \(\mathcal{U}_k\)'s is thus given by \cite{Lifang01EC} 
\begin{align}
&R_k^{\rm NOMA}(\nu)=\notag\\
&\left\{\begin{array}{l} \log_2\left(1+p_k(\nu)g_k(\nu)\right), \  {\rm if}\ g_k(\nu)>g_{\bar k}(\nu), \\
R_{k\rightarrow k}^{\rm NOMA}(\nu), \   {\rm otherwise.}
\end{array}\right. \label{eq:R_k NOMA under full CSIT}
\end{align}

Moreover, similar to \cite{Nihar03minimum}, we simultaneously consider two types of transmit power constraints on \(p_k\)'s, namely, average power constraint (APC) and peak power constraint (PPC), in which the former constrains the total transmit power in the long term, i.e., \(\mb{E}_\nu[p_k(\nu)+p_{\bar k}(\nu)]\le\bar P\), and the latter limits the instantaneous total transmit power below \(\hat P\), i.e., \(p_k(\nu)+p_{\bar k}(\nu)\le \hat P\), \(\forall\nu\). It is assumed that \(\bar P\le\hat P\) without loss of generality (w.l.o.g.). 
 
\subsubsection{OMA-Type-II}
For OMA-Type-II transmission, each user receives its information over \(\alpha_k(\nu)\) of the time/frequency dedicated to it in fading state $\nu$, such that \(\alpha_k(\nu)+\alpha_{\bar k}(\nu)=1\), where \(\alpha_k(\nu)\in[0,1]\). The same sets of transmit power constraints as in its NOMA counterpart, i.e., APC and PPC, are taken into account as well. Accordingly, the instantaneous achievable rate for \(\mathcal{U}_k\)'s in fading state \(\nu\) is given by\textcolor{black}{\footnote{If \(\alpha_k(\nu)=0\), we define \(R_k^{\rm OMA\text{-}II}(\nu)=0\), since \(\lim\limits_{\alpha_k\to 0^+}R_k^{\rm OMA\text{-}II}=0\).}}
\begin{align}
R_k^{\rm OMA\text{-}II}(\nu)=\alpha_k(\nu)\log_2\left(1+\frac{p_k(\nu)g_k(\nu)}{\alpha_k(\nu)}\right). \label{eq:R_k OMA-II under full CSIT}
\end{align}
Note that \eqref{eq:R_k OMA-II under full CSIT} applies to both TDMA and FDMA transmission in the sense that the total energy consumed for the two users in fading state \(\nu\) over time remains the same as that over frequency, which is given by \(\alpha_k(\nu)\frac{p_k(\nu)}{\alpha_k(\nu)}+\alpha_{\bar k}(\nu)\frac{p_{\bar k}(\nu)}{\alpha_{\bar k}(\nu)}=p_k(\nu)+p_{\bar k}(\nu)\), \(\forall\nu\).

\subsection{Partial CSIT}
\subsubsection{NOMA}
Under partial CSIT, for NOMA transmission, the BS does not know the exact CSI of the two users due to insufficient channel estimation but their relation, i.e., whether \(g_k(\nu)\le g_{\bar k}(\nu)\) or \(g_k(\nu)< g_{\bar k}(\nu)\), and statistical characteristics, \textcolor{black}{and therefore the Tx cannot dynamically adjust the allocation of power, rate and/or time/frequency resources to each fading state as in full CSIT}.  Hence, we adopt a binary power allocation strategy depending on which user has better CSI\textcolor{black}{\footnote{Such power policy under partial CSIT is not necessarily optimal but provided as performance lower-bound in comparison with its counterpart under full CSIT.}}. Specifically, in each fading state \(\nu\), an amount of power \(p_s\) is always assigned to the stronger user while \(p_w\) is assigned to the other weaker user. We also assume that \(p_s\) and \(p_w\) are static over all fading states, and therefore only APC applies, i.e., \(p_s+p_w\le\bar P\). In this case, the instantaneous rate \(R_k^{\prime\rm NOMA}(\nu)\)'s for \(\mathcal{U}_k\)'s is expressed as 
\begin{align}
  &R_k^{\prime\rm NOMA}(\nu)=\notag\\
  &\left\{\begin{array}{l} \log_2\left(1+p_sg_k(\nu)\right), \ {\rm if}\ g_k(\nu)>g_{\bar k}(\nu), \\
 \log_2\left(1+\frac{p_wg_k(\nu)}{p_sg_k(\nu)+1}\right), \  {\rm otherwise.}\end{array}\right.   \label{eq:R_k prime NOMA under partial CSIT}
\end{align} 
\subsubsection{OMA-Type-II}
Similarly, for OMA-Type-II transmission, the binary allocation policy with a fixed sharing of time/frequency between the two users is adopted. Specifically, the signal intended for \(\mathcal{U}_k\) is transmitted with power \(p_s\) if its channel gain from the BS is stronger than \(\mathcal{U}_{\bar k}\)'s, and with power \(p_w\) otherwise. The fixed proportion of time/frequency assigned to \(\mathcal{U}_k\) and \(\mathcal{U}_{\bar k}\) is \(\alpha_k\) and \(\alpha_{\bar k}\), respectively. Consequently, the instantaneous achievable rate for \(\mathcal{U}_k\)'s is expressed as
\begin{align}
&R_k^{\prime\rm OMA\text{-}II}(\nu)=\notag\\
&\left\{\begin{array}{l} \alpha_k\log_2\left(1+\frac{p_sg_k(\nu)}{\alpha_k}\right), \ {\rm if}\ g_k(\nu)>g_{\bar k}(\nu), \\
 \alpha_k\log_2\left(1+\frac{p_wg_k(\nu)}{\alpha_k}\right), \ {\rm otherwise.}\end{array}\right. \label{eq:R_k prime OMA-II under partial CSIT}
\end{align}

\subsection{System Throughput}
\subsubsection{Delay-Tolerant Transmission}
\textcolor{black}{First, for ``delay-tolerant'' transmission, we refer it to the scenario in which no delay constraints are imposed for decoding, and thus the codeword can be designed arbitrarily long (approaching infinity in theory ) spanning over all the fading states, and decoded until it is received in its full length. The associated performance metric for each user is {\em ergodic rate} \cite{Lifang01EC}, at which the Tx delivers the intended data for each user over the entire fading process.}  Consequently, the ergodic sum-rate (ESR) of the two users are given by \(\mb{E}_\nu[R_k^{\rm NOMA}(\nu)+R_{\bar k}^{\rm NOMA}(\nu)]\) (\(\mb{E}_\nu[R_k^{\prime\rm NOMA}(\nu)+R_{\bar k}^{\prime\rm NOMA}(\nu)]\)), and \(\mb{E}_\nu[R_k^{\rm OMA\text{-}II}(\nu)+R_{\bar k}^{\rm OMA\text{-}II}(\nu)]\) (\(\mb{E}_\nu[R_k^{\prime\rm OMA\text{-}II}(\nu)+R_{\bar k}^{\prime\rm OMA\text{-}II}(\nu)]\)), for NOMA and OMA-Type-II transmission, respectively\footnote{Since the analysis developed for ``delay-tolerant'' transmission may also apply to scenarios, in which the short-length codewords are detected at each user on a block basis and the average sum-rate is used to measure the achievable sum-rate in the long term, we do not explicitly differentiate the two terms, ``ESR'' and ``average sum-rate'', throughout the paper.}. 
 
\subsubsection{Delay-Limited Transmission}
Next, consider the delay-limited types of transmission for downlink NOMA and/or OMA-Type-II system. \textcolor{black}{We relax the classical information theoretic ``zero-outage'' definition in \cite{Stephen98delay-limited}. Other than maintain a constant rate vector at all fading states via power control, we refer this notion to the scenario in which delay-sensitive data such as video streaming requires to be correctly decoded at a constant rate at the end of every fading state\textcolor{black}{\footnote{We assume in the ``delay-limited'' transmission that SIC can be perfectly performed during one block, which is hardly true in practice and thus provides theoretical upper-bound for the achievable sum of DLT. This assumption may be lifted by explicitly considering imperfect SIC as in \cite{Hina17ModelingPCP} in our future work.}}. The associated performance metric for each user is {\em outage probability}, which measures the percentage of fading states at which a predefined constant rate cannot be supported.} 

Specifically, under full CSIT, the outage probability for user \(\mathcal{U}_k\) with the target rate \(\bar R_k\), \(k\in\{1,2\}\), is introduced as below.\\
{\bf Case 1}: \(g_k(\nu)>g_{\bar k}(\nu)\)
\begin{multline}
\zeta_k^{\rm NOMA}=\Pr\left\{R_{k\rightarrow \bar{k}}^{\rm NOMA}(\nu)<\bar R_{\bar k}, R_{k\rightarrow k}^{\rm NOMA}(\nu)<\bar R_k\right\}+\\
\Pr\left\{R_{k\rightarrow \bar{k}}^{\rm NOMA}(\nu)\ge\bar R_{\bar k}, R_k^{\rm NOMA}(\nu)<\bar R_k\right\}. \label{eq:zeta_k NOMA for strong user under full CSIT}
\end{multline}
{\bf Case 2}: \(g_k(\nu)\le g_{\bar k}(\nu)\)
\begin{align}
\zeta_k^{\rm NOMA}=\Pr\left\{R_k^{\rm NOMA}(\nu)<\bar R_k\right\}. \label{eq:zeta_k NOMA for weak user under full CSIT}
\end{align} 
As seen from \eqref{eq:zeta_k NOMA for strong user under full CSIT}, when  \(\mathcal{U}_k\) has better channel condition, whether its signal-to-noise ratio (SNR) or signal-to-interference-plus-noise ratio (SINR) leads to its outage depends on whether or not it manages to recover \(\mathcal{U}_{\bar k}\)'s message. If it fails to retrieve \(\mathcal{U}_{\bar k}\)'s message at the predefined transmission rate for \(\mathcal{U}_{\bar k}\), i.e., \(\bar R_{\bar k}\), it has to decode its own by TIAN. Otherwise, if it succeeds in decoding \(\mathcal{U}_{\bar k}\)'s message, SIC is performed before it decodes its own interference-free. On the other hand, when \(\mathcal{U}_k\) has worse channel condition, it always decodes its own by TIAN (c.f.~\eqref{eq:zeta_k NOMA for weak user under full CSIT}). 

In addition, at each fading state \(\nu\), an outage indicator function is defined as follows \cite{xing2016secrecySWIPT}.\\
{\bf Case 1}: \(g_k(\nu)>g_{\bar k}(\nu)\)
\begin{align}
&X_k^{\rm NOMA}(\nu)=\notag\\
&\left\{\begin{array}{ll} 1,& {\rm if}\ R_{k\rightarrow\bar{k}}^{\rm NOMA}(\nu)<\bar R_{\bar k}, \ R_{k\rightarrow k}^{\rm NOMA}(\nu)<\bar R_k,\\
1, & {\rm if}\ R_{k\rightarrow\bar{k}}^{\rm NOMA}(\nu)\ge\bar R_{\bar k}, \ R_k^{\rm NOMA}(\nu)<\bar R_k, \\
0, & {\rm otherwise.}
\end{array}\right. \label{eq:X_k NOMA for strong user under full CSIT}
\end{align} 
{\bf Case 2}: \(g_k(\nu)\le g_{\bar k}(\nu)\)
\begin{align}
X_k^{\rm NOMA}(\nu)=
\left\{\begin{array}{ll} 1, & {\rm if}\ \ R_{k\rightarrow k}^{\rm NOMA}(\nu)<\bar R_k,\\
0, &{\rm otherwise.}
\end{array}\right. \label{eq:X_k NOMA for weak user under full CSIT}
\end{align} 
Combining \eqref{eq:X_k NOMA for strong user under full CSIT} (c.f.~\eqref{eq:zeta_k NOMA for strong user under full CSIT}) and \eqref{eq:X_k NOMA for weak user under full CSIT} (c.f.~\eqref{eq:zeta_k NOMA for weak user under full CSIT}), it is easily verified that \(\mb E_\nu[X_k^{\rm NOMA}(\nu)]=\zeta_k^{\rm NOMA}\), \(k\in\{1,2\}\).
  
For OMA-Type-II transmission, the outage probability of \(\mathcal{U}_k\)'s is defined independent of the other as follows:
\begin{align}
\zeta_k^{\rm OMA\text{-}II}=\Pr\left\{R_k^{\rm OMA\text{-}II}(\nu)<\bar R_k\right\}. \label{eq:sigma_k OMA-II under full CSIT}
\end{align}
By analogy, we introduce the following indicator function for \(\mathcal{U}_k\) w.r.t. the target rate \(\bar R_k\):
 \begin{align}
 X_k^{\rm OMA\text{-}II}(\nu)=\left\{\begin{array}{ll} 1, & {\rm if}\ R_k^{\rm OMA\text{-}II}(\nu)<\bar R_k,\\
0, &{\rm otherwise.}
\end{array}\right. \label{eq:X_k OMA-II under full CSIT}
 \end{align} It also follows that \(\mb E_\nu[X_k^{\rm OMA\text{-}II}(\nu)]=\zeta_k^{\rm OMA\text{-}II}\), \(k\in\{1,2\}\).
 
Accordingly, one relevant metric to assess the overall performance in delay-limited case is the {\em sum of DLT} expressed as \(\bar R_k(1-\zeta_k^{\rm NOMA})+\bar R_{\bar k}(1-\zeta_{\bar k}^{\rm NOMA})\), and \(\bar R_k(1-\zeta_k^{\rm OMA\text{-}II})+\bar R_{\bar k}(1-\zeta_{\bar k}^{\rm OMA\text{-}II})\), for NOMA and OMA-Type-II transmission under full CSIT, respectively. The sum of DLT for NOMA and OMA-Type-II transmission under partial CSIT is also similarly given by \(\bar R_k(1-\zeta_k^{\prime\rm NOMA})+\bar R_{\bar k}(1-\zeta_{\bar k}^{\prime\rm NOMA})\), and \(\bar R_k(1-\zeta_k^{\prime\rm OMA\text{-}II})+\bar R_{\bar k}(1-\zeta_{\bar k}^{\prime\rm OMA\text{-}II})\), respectively.

\section{Optimum Delay-Tolerant Transmission}\label{sec:Optimum Delay-Tolerant Transmission}
In delay-tolerant scenarios, to maximize the ESR of the system while guaranteeing certain level of fairness, a minimum achievable ergodic rate requirement for each user, namely, \(\mb{E}_\nu[R_k^{\rm XX}(\nu)]\ge\bar R\) (\(\mb{E}_\nu[R_k^{\prime\rm XX}(\nu)]\ge\bar R^\prime\)), \(k\in\{1,2\}\), is imposed, where \((\cdot)^{\rm XX}\) denotes the multiple access scheme that is specified in the context throughout the paper. In this section, the optimal trade-off between the system ESR and user fairness is pursued in the case of full and partial CSIT, respectively. \textcolor{black}{Particularly, under full CSIT, the ESR maximization problems are solved using Lagrangian dual decomposition levering ``time-sharing'' conditions, while under partial CSIT, individual user's ergodic rate needs to be first analysed in closed form by means of CDFs of the related SNR and/or SNRs.}
 
\subsection{Full CSIT}
In the case of full CSIT, the design objective is to maximize the system ESR by jointly optimizing the power and/or orthogonal resource allocations, and the two users' instantaneous rate at each fading state, subject to both APC and PPC at the BS, as well as a minimum ergodic rate constraint for the two users. As a result, the optimization problem is formulated as follows\footnote{Note that \(\mb E[\cdot]\) in (P1-XX) is evaluated by the sum of the associated instantaneous function of \(\nu\) divided by the total number of fading states $N$, assuming that $N$ is large enough such that \(N\to\infty\).}.  
\begin{subequations}  
\begin{align}
\mathrm{(P1\text{-}XX)}:&~~\mathop{\mathtt{Maximize}}_{\{p_k(\nu), p_{\bar k}(\nu),\alpha_k(\nu)\}}\mb{E}_\nu[R_k^{\rm XX}(\nu)+R_{\bar k}^{\rm XX}(\nu)]\notag\\
&~~\mathtt{Subject \ to}\notag\\
&~~\mb{E}_\nu[p_k(\nu)+p_{\bar k}(\nu)]\leq \bar P,\label{eq:APC under full CSIT}\\
&~~p_k(\nu)+p_{\bar k}(\nu)\leq \hat P, \ \forall\nu,\\
&~~p_k(\nu)\ge 0, \ p_{\bar k}(\nu)\ge 0,\ \forall\nu,\\
&~~\mb{E}_\nu[R_k^{\rm XX}(\nu)]\ge\bar R, \; \forall k,\label{eq:min rate constraint under full CSIT}
\end{align}
\end{subequations} 
where the exclusive parameters for OMA-Type-II, \(\{\alpha_k(\nu)\}\)'s, are only valid when \({\rm XX}\) refers to OMA-Type-II. In the following, we develop optimal solution to \(\mathrm{(P1\text{-}NOMA)}\) and \(\mathrm{(P1\text{-}OMA\text{-}II)}\), respectively.
\subsubsection{Optimal Solution to \(\mathrm{(P1\text{-}NOMA)}\)}\label{subsubsec:Optimal Solution to (P1-NOMA) under Full CSIT}
Problem \(\mathrm{(P1\text{-}NOMA)}\) is non-convex due to the non-convex objective function (c.f. \eqref{eq:R_k NOMA under full CSIT}), and therefore no immediate solution can be given. However, for channel fading following continuous distributions, \(\mathrm{(P1\text{-}NOMA)}\) proves to satisfy the ``time-sharing''  condition\textcolor{black}{\footnote{The original definition of ``time-sharing'' condition is given by \cite[Definition 1]{Yu2006}, which essentially implies that the maximum value of the optimization problem (P1-NOMA) is a joint concave function of \(\bar P\) and \(\bar R\). The proof is  rather  standard and thus omitted herein for brevity. }}. \textcolor{black}{Note that if (P1-NOMA) satisfies the ``time-sharing condition'', then it has a zero duality gap between the primal and the dual problem using Lagrangian duality \cite[Theorem 1]{Yu2006}, i.e., {\em strong duality} \cite{rockafellar1997convex} holds, despite of the convexity of the problem itself.} Hence, we can still optimally solve it via its dual problem.
 
Next, we apply the Lagrangian dual method to solve \(\mathrm{(P1\text{-}NOMA)}\),  the  Lagrangian of which is given by
\begin{multline}
\mathcal{L}_1^{\rm NOMA}(\{p_k(\nu)\},\{p_{\bar k}(\nu)\},\lambda,\delta,\mu)=\\
%=\mb{E}_\nu[R_k^{\rm NOMA}(\nu)+R_{\bar k}^{\rm NOMA}(\nu)]-\lambda(\mb{E}_\nu[p_k(\nu)+p_{\bar k}(\nu)]-\bar P)\\
%+\delta(\mb{E}_\nu[R_k(\nu)]-\bar R)+\mu(\mb{E}_\nu[R_{\bar k}(\nu)]-\bar R)\\
\mb{E}_\nu[(1+\delta)R_k^{\rm NOMA}(\nu)+(1+\mu)R_{\bar k}^{\rm NOMA}(\nu)-\lambda(p_k(\nu)+p_{\bar k}(\nu))]\\
+\lambda\bar P-\delta\bar R-\mu\bar R, \label{eq:Lagrangian of (P1-NOMA)}
\end{multline}
where \(\lambda\) is the Lagrangian multiplier associated with the APC given in \eqref{eq:APC under full CSIT}; \(\delta\) and \(\mu\) are those associated with the ergodic rate constraints given in \eqref{eq:min rate constraint under full CSIT} for \(\mathcal{U}_k\) and \(\mathcal{U}_{\bar k}\), respectively. The dual function of \(\mathrm{(P1\text{-}NOMA)}\) corresponding to \eqref{eq:Lagrangian of (P1-NOMA)} is accordingly given by
\begin{align}
&g(\lambda,\delta,\mu)=\max
\mathcal{L}_1^{\rm NOMA}(\{p_k(\nu)\},\{p_{\bar k}(\nu)\},\lambda,\delta,\mu),\notag\\
&\mathtt{s.t.}~~p_k(\nu)\ge 0, p_{\bar k}(\nu)\ge 0, p_k(\nu)+p_{\bar k}(\nu)\le\hat P, \; \forall\nu. \label{eq:dual function of (P1)}
\end{align}
The dual problem of \(\mathrm{(P1\text{-}NOMA)}\) is thus formulated as
\begin{align*}\mathrm{(P1\text{-}NOMA\text{-}dual)}:~\mathop{\mathtt{Minimize}}_{\lambda\ge 0,\delta\ge 0,\mu\ge 0}
& ~~~g(\lambda,\delta,\mu).
\end{align*} 

It is observed that \(g(\lambda,\delta,\mu)\) is obtained by maximizing the Lagrangian given in \eqref{eq:Lagrangian of (P1-NOMA)}, which can be decoupled into as many subproblems as the number of fading states all sharing the same structure. The index \(\nu\) is now safely dropped for the ease of exposition. Taking one particular fading state as an example, the associated subproblem given a triple \((\lambda,\delta,\mu)\) can be expressed as  
\begin{align*}
\mathrm{(P1\text{-}NOMA\text{-}sub)}:~\mathop{\mathtt{Maximize}}_{p_k\ge 0,p_{\bar k}\ge 0}
& ~~~\bar{\mathcal{L}}_1^{\rm NOMA}(p_k,p_{\bar k})\\
\mathtt {Subject \ to}& ~~~p_k+p_{\bar k}\leq\hat P,
\end{align*} where \(\bar{\mathcal{L}_1}^{\rm NOMA}(p_k,p_{\bar k})=(1+\delta)R_k^{\rm NOMA}+(1+\mu)R_{\bar k}^{\rm NOMA}-\lambda(p_k+p_{\bar k})\).
Since these problems are independent of each other, they can be solved in parallel each for one fading state. Therefore, w.l.o.g., we focus on solving \(\mathrm{(P1\text{-}NOMA\text{-}sub)}\) in the sequel.

\begin{proposition}
The optimal power allocation to Problem \(\mathrm{(P1\text{-}NOMA\text{-}sub)}\) assuming \(g_1>g_2\) is given by\textcolor{black}{\footnote{In fact, under full CSIT, following the same method utilized to develop Proposition~\ref{prop:optimal power allocation for ESR NOMA}, the two-user results can be generalized to more general cases with $K>2$, since the difficulty of solving \eqref{eq:stationary point} does not increase with $K$.}}
\begin{multline}
(p_1^\ast, p_2^\ast)=\arg\max\{\bar{\mathcal{L}}_1^{\rm NOMA}(0,0),\bar{\mathcal{L}}_1^{\rm NOMA}(0,\hat P),\\
\bar{\mathcal{L}}_1^{\rm NOMA}(\hat P,0), \bar{\mathcal{L}}_1^{\rm NOMA}(p_{i,1},p_{i,2})\}, \;   i\in\{1,2,3,4\},
\end{multline} \textcolor{black}{where \((p_{i,1},p_{i,2})\), \(i\in\{1,2,3,4\}\), are given at the top of the next page with each corresponding to one solution pair given by \eqref{eq:case-specific optimal power allocation for ESR NOMA under full CSIT}.} \label{prop:optimal power allocation for ESR NOMA}
\end{proposition}
\begin{IEEEproof}
Since \(\bar{\mathcal{L}}_1^{\rm NOMA}(p_1,p_2)\) is a continuous function over \(\Psi=\{(p_1,p_2)|p_1\ge 0, p_2\ge 0, p_1+p_2\le\hat P\}\), its maximum proves to be either at the stationary point, denoted by \((p_{4,1},p_{4,2})\), or on the boundary of \(\Psi\) depending on whether \((p_{4,1},p_{4,2})\in\Psi\) or not. We calculate \((p_{4,1},p_{4,2})\) as follows:
\begin{align}
(p_{4,1},p_{4,2})=\arg\left\{\nabla_{(p_1,p_2)}\bar{\mathcal{L}}_1^{\rm NOMA}(p_1,p_2)=\mv 0\right\}. \label{eq:stationary point}
\end{align} 
If \((p_{4,1},p_{4,2})\in\Psi\), the maximum is \(\bar{\mathcal{L}}_1^{\rm NOMA}(p_{4,1},p_{4,2})\), otherwise the maximum can be attained by restricting \((p_1,p_2)\) to the lines \(p_1=0\), \(p_2=0\) or \(p_1+p_2=\hat P\). The stationary points on these lines are denoted by \((p_{i,1},p_{i,2})\)'s, \(i=1\), \(2\) and \(3\), respectively.
\end{IEEEproof}

Note that Proposition \ref{prop:optimal power allocation for ESR NOMA} assumes \(g_1>g_2\) for the ease of exposition though, its results also apply to the fading states where \(g_1<g_2\) by simply exchanging \(\delta\), \(p_1\), and \(g_1\) with \(\mu\), \(p_2\), and \(g_2\), respectively, in \eqref{eq:case-specific optimal power allocation for ESR NOMA under full CSIT}.  \textcolor{black}{Some optimal system design insights are gained from Proposition~\ref{prop:optimal power allocation for ESR NOMA}. 
%First, assuming the optimal dual variables denoted by \((\lambda^\ast,\delta^\ast,\mu^\ast)\) and the optimum average user rate satisfies \(\mb{E}_\nu[R_1^{\rm NOMA}(\nu)]>\bar R \) and \(\mb{E}_\nu[R_2^{\rm NOMA}(\nu)]=\bar R\), in accordance with complementary slackness in Karush-Kuhn-Tucker  (KKT) conditions, it follows that \(\delta^\ast=0\). In this case \((p_{3,1},p_{3,2})\) reduces to 
%\begin{align}
%p_{3,1} & =\left[\hat P+\frac{(\hat P+\Myfrac{1}{g_2})-(1+\mu^\ast)(\hat P+\Myfrac{1}{g_1})}{\mu^\ast}\right]_0^{\hat P}, \label{eq:reduced expression of p_{3,1}}\\   
%p_{3,2} & =\left[\frac{(1+\mu^\ast)(\hat P+\Myfrac{1}{g_1})-(\hat P+\Myfrac{1}{g_2})}{\mu^\ast}\right]_0^{\hat P}. \label{eq:reduced expression of p_{3,2}}
%\end{align}
%It is thus observed from \eqref{eq:reduced expression of p_{3,1}} and \eqref{eq:reduced expression of p_{3,2}} that when \((p_{1}^\ast,p_{2}^\ast)=(p_{3,1},p_{3,2})\), if \(\mu^\ast>\frac{\hat P+\Myfrac{1}{g_2}}{\hat P+\Myfrac{1}{g_1}}-1\), the weak user \(\mathcal{U}_2\) starts the transmission due to the average rate requirement of \(\mb{E}_\nu[R_2^{\rm NOMA}(\nu)]\ge\bar R\); otherwise all the instantaneous power \(\hat P\) is allocated to \(\mathcal{U}_1\) for the purpose of maximizing \(\bar{\mathcal{L}}_1^{\rm NOMA}(p_1,p_2)\). Furthermore, 
Considering an extreme case in favour of \(\mathcal{U}_2\), in which \(\delta\ll\mu\), it is observed from \eqref{eq:case-specific optimal power allocation for ESR NOMA under full CSIT} that \(p_{4,1}\) monotonically decreases with \(\mu\) while \(p_{4,2}\) monotonically increases with \(\mu\), which suggests that when \(\mu\) associated with \(\mathcal{U}_2\)'s QoS requirement is sufficiently large, the optimal power allocation policy tends to suppress \(\mathcal{U}_1\)'s transmission while supporting \(\mathcal{U}_2\)'s despite of \(\mathcal{U}_1\)'s channel condition better than \(\mathcal{U}_2\).} 

\begin{figure*}
	{\small\begin{align}
		\kern-8pt\left\{\begin{array}{ll} 
		p_{1,1}=0, & p_{1,2}=\left[\frac{1+\mu}{\lambda\ln 2}-\frac{1}{g_2}\right]_0^{\hat P}\\ 
		p_{2,1}=\left[\frac{1+\delta}{\lambda\ln 2}-\frac{1}{g_1}\right]_0^{\hat P}, & p_{2,2}=0\\
		p_{3,1}=\left[\frac{\Myfrac{(1+\mu)}{g_1}-\Myfrac{(1+\delta)}{g_2}}{\delta-\mu}\right]_0^{\hat P}, & p_{3,2}=\left[\hat P-\frac{\Myfrac{(1+\mu)}{g_1}-\Myfrac{(1+\delta)}{g_2}}{\delta-\mu}\right]_0^{\hat P}\\
		\left\{\begin{array}{l}
		p_{4,1}=\frac{\Myfrac{(1+\mu)}{g_1}-\Myfrac{(1+\delta)}{g_2}}{\delta-\mu}\\
		p_{4,2}=\frac{1+\mu}{\lambda\ln 2}-\frac{1}{g_2}-\frac{\Myfrac{(1+\mu)}{g_1}-\Myfrac{(1+\delta)}{g_2}}{\delta-\mu}
		\end{array},\right. &\left.\begin{array}{l}
		\mbox{if}\ p_{4,1}\ge 0, p_{4,2}\ge 0, p_{4,1}+p_{4,2}\le\hat P,\\
		\mbox{N/A, otherwise.}\end{array}\right.
		\end{array}\right.\label{eq:case-specific optimal power allocation for ESR NOMA under full CSIT}
		\end{align}}
	\hrulefill
\end{figure*}

Thanks to Proposition~\ref{prop:optimal power allocation for ESR NOMA}, given a triple \((\lambda,\delta,\mu)\), \(g(\lambda,\delta,\mu)\) is obtained efficiently by solving \(\mathrm{(P1\text{-}NOMA\text{-}sub)}\) in parallel over all fading states. \(\mathrm{(P1\text{-}NOMA\text{-}dual)}\) can thus be iteratively solved using sub-gradient based methods, e.g., deep-cut ellipsoid method (with constraints) \cite[Localization methods]{EE364b}.
%whose objective iteration is briefly given by
%\begin{align}
%\mv x^{(i+1)}=\mv x^{(i)}-\frac{1+3\alpha^{(i+1)}}{4}\mv P^{(i)}\tilde{\mv s}^{(i+1)}, \label{eq:dual variable update}
%\end{align}
%where \(i\) denotes the \(i\)th iteration, \(\mv x=[\lambda,\delta,\mu]^T\), \(\mv s\) is the sub-gradient w.r.t. the triple \((\lambda,\delta,\mu)\), and \(\tilde{\mv s}\) is \(\mv s\) normalized by \(\sqrt{\mv s^T\mv P\mv s}\). In addition, \(\alpha\) and \(\mv P\) are also parameters related to ellipsoid method, 
The required sub-gradient for updating \((\lambda,\delta,\mu)\) turns out to be \((\bar P-\mb{E}_\nu[p_k^\ast(\nu)+p_{\bar k}^\ast(\nu)], \mb{E}_\nu[R_k^{\ast\rm NOMA}(\nu)]-\bar R, \mb{E}_\nu[R_{\bar k}^{\ast\rm NOMA}(\nu)]-\bar R)^T\), where \((p_k^\ast(\nu),p_{\bar k}^\ast(\nu))\) is the optimal solution to \(\mathrm{(P1\text{-}NOMA\text{-}sub)}\) at fading state \(\nu\), and \(R_k^{\ast\rm NOMA}(\nu)\)'s is obtained by substituting \((p_k^\ast(\nu),p_{\bar k}^\ast(\nu))\) into \eqref{eq:R_k NOMA under full CSIT}.

%{\small\begin{table}[htp]
%\begin{center}
%\caption{}\label{table:Algorithm I}
%\vspace{-0.75em}
%\hrule
%\vspace{0.50em}
%\begin{algorithmic}[1]
%\REQUIRE \(\lambda^{(0)}\), \(\delta^{(0)}\), \(\mu^{(0)}\), \(\alpha^{(0)}\), and \(\mv P^{(0)}\)
%\REPEAT
%\STATE solve \(\mathrm{(P1\text{-}NOMA\text{-}sub)}\) given \((\lambda^{(i)},\delta^{(i)},\mu^{(i)})\) according to Proposition~\ref{prop:optimal power allocation for ESR NOMA} and obtain \(\{p_k(\nu), p_{\bar k}(\nu)\}\)
%\STATE update \(\mv s^{(i)}=[\bar P-\mb{E}_\nu[p_k(\nu)+p_{\bar k}(\nu)], \mb{E}_\nu[R_k^{\rm NOMA}(\nu)]-\bar R, \mb{E}_\nu[R_{\bar k}^{\rm NOMA}(\nu)]-\bar R]^T\),  \(\alpha^{(i)}\) and \(\tilde{\mv s}^{(i)}\)
%\STATE update \(\mv x^{(i)}\) according to \eqref{eq:dual variable update}
%\STATE update \(\mv P^{(i)}\)
%\UNTIL \(\sqrt{\mv s^T\mv P\mv s}\le\epsilon_{\rm obj}\)\\
%\ENSURE \(\{p_k^\ast(\nu), p_{\bar k}^\ast(\nu)\}\leftarrow\{p_k(\nu), p_{\bar k}(\nu)\}\)
%\end{algorithmic}
%\vspace{0.50em}
%\hrule
%\end{center}
%\vspace{-1.0em}
%\end{table}}

Note that a feasible \(\bar R\) in \eqref{eq:min rate constraint under full CSIT} ensures the successful implementation of the ellipsoid method, and thus it is important to consider a reasonable \(\bar R\) that does not exceed \(\bar R_{\max}\). We can obtain \(\bar R_{\max}\) by replacing the objective function of \(\mathrm{(P1\text{-}NOMA)}\) with a variable \(\bar R\) and then solving the feasibility problem by bi-section over \(\bar R\). Since the involved procedure is quite similar to that for solving \(\mathrm{(P1\text{-}NOMA)}\), we omit it herein for brevity.
\subsubsection{Optimal Solution to \(\mathrm{(P1\text{-}OMA\text{-}II)}\)}\label{subsubsec:Optimal Solution to (P1-OMA-II) under Full CSIT}
First, \(\mathrm{(P1\text{-}OMA\text{-}II)}\) is a convex problem, since \eqref{eq:R_k OMA-II under full CSIT} as the perspective of the jointly concave function \(\log_2(1+p_k(\nu)g_k(\nu))\) proves to be jointly concave w.r.t. \(\alpha_k(\nu)\) and \(p_k(\nu)\), \(k\in\{1,2\}\), \(\forall\nu\). As such, we can solicit the Lagrangian dual method to solve \(\mathrm{(P1\text{-}OMA\text{-}II)}\) due to strong duality. 

The Lagrangian of \(\mathrm{(P1\text{-}OMA\text{-}II)}\) is expressed as
\begin{multline}
\mathcal{L}_1^{\rm OMA\text{-}II}(\{p_k(\nu)\},\{p_{\bar k}(\nu)\},\{\alpha_k(\nu)\},\lambda,\delta,\mu)=\\
\mb{E}_\nu[(1+\delta)R_k^{\rm OMA\text{-}II}(\nu)+(1+\mu)R_{\bar k}^{\rm OMA\text{-}II}(\nu)-\lambda(p_k(\nu)+p_{\bar k}(\nu))]\\
+\lambda\bar P-\delta\bar R-\mu\bar R, \label{eq:Lagrangian of (P1-OMA-II)}
\end{multline} where \(\lambda\), \(\delta\) and \(\mu\) are Lagrangian multipliers associated with the same constraints as those for \(\mathrm{(P1\text{-}NOMA)}\). Similar to the previous section, \(\mathcal{L}_1^{\rm OMA\text{-}II}(\{p_k(\nu)\},\{p_{\bar k}(\nu)\},\{\alpha_k(\nu)\},\lambda,\delta,\mu)\) can also be decoupled into parallel sub-Lagrangian all having the same structure. We define \(\bar{\mathcal{L}}_1^{\rm OMA\text{-}II}(p_k,p_{\bar k},\alpha_k)=(1+\delta)R_k^{\rm OMA\text{-}II}+(1+\mu)R_{\bar k}^{\rm OMA\text{-}II}-\lambda(p_k+p_{\bar k})\). Then the associated subproblem one particular fading state is formulated as
\begin{align*}
\mathrm{(P1\text{-}OMA\text{-}II\text{-}sub)}:~\mathop{\mathtt{Maximize}}_{p_k\ge 0,p_{\bar k}\ge 0,\alpha_k}
& ~~~\bar{\mathcal{L}}_1^{\rm OMA\text{-}II}(p_k,p_{\bar k},\alpha_k)\\
\mathtt {Subject \ to}& ~~~p_k+p_{\bar k}\leq\hat P,\\
&~~~0\le\alpha_k\le 1, \; \forall k,
\end{align*} where the index \(\nu\) has been dropped for the ease of exposition. To solve \(\mathrm{(P1\text{-}OMA\text{-}II\text{-}sub)}\), the following two lemmas are required.

\begin{lemma}
If the maximum of \(\bar{\mathcal{L}}_1^{\rm OMA\text{-}II}(p_1,p_2,\alpha_1)\) is achieved by its jointly stationary point, it is necessary to have the following conditions satisfied:
\begin{subequations}
\begin{align}
& h(\lambda,\delta,\mu)=0,\label{eq:h(lambda,delta,mu)=0}\\
& c_1\ge 0,\label{eq:c1 non-negative}\\
& c_2\ge 0,\label{eq:c2 non-negative}\\
&\left\{\begin{array}{ll} \frac{\hat P-c_2}{c_1-c_2}\ge 0, & {\rm if}\ c_1>c_2,\\
\frac{\hat P-c_2}{c_1-c_2}\le 1, &{\rm otherwise,}
\end{array},\right.\label{eq:necessary conditions for sp in OMA-II}
\end{align}
\end{subequations} where \(c_1=\frac{1+\delta}{\lambda\ln 2}-\frac{1}{g_1}\), \(c_2=\frac{1+\mu}{\lambda\ln 2}-\frac{1}{g_2}\), and \(h(\lambda,\delta,\mu)\) is given by
\begin{multline}
h(\lambda,\delta,\mu)=(1+\delta)\log_2\left(\frac{1+\delta}{\lambda\ln 2}g_1 \right)-\\
(1+\mu)\log_2\left(\frac{1+\mu}{\lambda\ln 2}g_2 \right )-\lambda c_1+\lambda c_2. \label{eq:h(lambda,delta,mu)}
\end{multline} The corresponding stationary point is given by
\begin{align}
p_1^\ast=c_1\alpha_1^\ast, \ p_2^\ast=c_2(1-\alpha_1^\ast), \label{eq:sp in OMA-II} 
\end{align} where
\begin{align}
\alpha_1^\ast=\left\{\begin{array}{ll}
\forall\in[0,\min\{\frac{\hat P-c_2}{c_1-c_2},1\}], &{\rm if}\ c_1>c_2,\\
\forall\in[(\frac{\hat P-c_2}{c_1-c_2})^+,1], & {\rm otherwise.}\end{array}\right. \label{eq:sp of alpha in OMA-II} 
\end{align}\label{lemma:sp in OMA-II}
\end{lemma}
\begin{IEEEproof}
First, solve \(\nabla_{(p_1,p_2,\alpha_1)}\bar{\mathcal{L}}_1^{\rm OMA\text{-}II}(p_1,p_2,\alpha_1)=\mv 0\) to obtain the jointly stationary point. Next, by plugging \(p_1=c_1\alpha_1\) and \(p_2=c_2(1-\alpha_1)\) into the partial derivative of \(\bar{\mathcal{L}}_1^{\rm OMA\text{-}II}(p_1,p_2,\alpha_1)\) w.r.t. \(\alpha_1\), \eqref{eq:h(lambda,delta,mu)=0} is obtained. Finally, constrain \(P_1\ge 0\), \(p_2\ge 0\), and \(p_1+p_2\le\hat P\), we arrive at \eqref{eq:c1 non-negative}, \eqref{eq:c2 non-negative}, and the feasible range for \(\alpha_1\) given in \eqref{eq:sp of alpha in OMA-II}, respectively.
\end{IEEEproof}
\begin{lemma}
If the maximum of \(\bar{\mathcal{L}}_1^{\rm OMA\text{-}II}(p_1,p_2,\alpha_1)\) is achieved by points on the boundary \(p_1+p_2=\hat P\), the optimum \((p_1,p_2,\alpha_1)\) turns out to be 
\begin{align}
\left\{\begin{array}{ll}
p_1^\ast=0, \ p_2^\ast=\hat P, \ \alpha_1^\ast=0, & {\rm if}\ \frac{1+\mu}{1+\delta}>\frac{\log_2(1+\hat Pg_1)}{\log_2(1+\hat Pg_2)},\\
p_1^\ast=\hat P, \ p_2^\ast=0, \ \alpha_1^\ast=1, & {\rm otherwise.}
\end{array}\right.  \label{eq:p1+p2=hat P in OMA-II}
\end{align} \label{lemma:p1+p2=hat P in OMA-II}
\end{lemma}  
\begin{IEEEproof}
Please refer to Appendix~\ref{appendix:proof of p1+p2=hat P in OMA-II}.
\end{IEEEproof}
Based on Lemma~\ref{lemma:sp in OMA-II} and Lemma~\ref{lemma:p1+p2=hat P in OMA-II}, the following proposition is derived.

\begin{proposition}
The optimal power as well as time/frequency allocation to \(\mathrm{(P1\text{-}OMA\text{-}II\text{-}sub)}\) is given by
\begin{multline}
(p_1^\ast, p_2^\ast,\alpha_1^\ast)=\arg\max\{\bar{\mathcal{L}}_1^{\rm OMA\text{-}II}(0,0,0),\\
\bar{\mathcal{L}}_1^{\rm OMA\text{-}II}(0,\hat P,0), \bar{\mathcal{L}}_1^{\rm OMA\text{-}II}(\hat P,0,1),\\ \bar{\mathcal{L}}_1^{\rm OMA\text{-}II}(0,c_2,0)\mathbbm{1}_{c_2},\bar{\mathcal{L}}_1^{\rm OMA\text{-}II}(c_1,0,1)\mathbbm{1}_{c_1}\}, \label{eq:optimal power allocation for ESR OMA-II}
\end{multline} where \(\mathbbm{1}_{(\cdot)}\) is an indicator function defined as
\begin{align}
\mathbbm{1}_x=\left\{\begin{array}{ll}
1, & {\rm if}\ 0\le x\le\hat P,\\
0, &{\rm otherwise.}\end{array}\right. \label{eq:power constraint indicator}
\end{align}\label{prop:optimal power allocation for ESR OMA-II}
\end{proposition}
\begin{IEEEproof}
Please refer to Appendix~\ref{appendix:proof of optimal power allocation for ESR OMA-II}.
\end{IEEEproof}

\begin{remark}
When there are only two users, the optimal solution given by \eqref{eq:optimal power allocation for ESR OMA-II} shares some philosophy in common with that achieves the boundary of the time division (TD) capacity region discussed in \cite[\emph{Theorem 3}]{Lifang01EC}. We focus on solving \(\mathrm{(P1\text{-}OMA\text{-}II\text{-}sub)}\) in any fading state given a triple \((\lambda,\delta,\mu)\), while \cite{Lifang01EC} maximized the total weighted sum-rate in any fading state by determining how to distribute \(P(\mv n)\) among \(M=2\) users such that the instantaneous total power constraint \(\sum_{j=1}^2\tau_jP_j(\mv n)=P(\mv n)\) (c.f.~\cite[Eqn.~(11)]{Lifang01EC}) is satisfied. The optimal solutions both suggest that with probability $1$, at most one single user transmits in any fading state. This is because the probability measure of any subset of \(\{(g_k(\nu),g_{\bar k}(\nu)):h(\lambda,\delta,\mu)=0\}\) (c.f.~\eqref{eq:h(lambda,delta,mu)}) assuming continuously joint distribution of \((g_k(\nu),g_{\bar k}(\nu))\) is zero, as is the probability measure of \(\{\mv n:h(\lambda,\mv n)=0\}\) in \cite[\emph{Theorem 3}]{Lifang01EC}. This also explains why the maximum of \(\bar{\mathcal{L}}_1^{\rm OMA\text{-}II}(p_1,p_2,\alpha_1)\) cannot be achieved by its jointly stationary point in probability. \label{remark:at most one user transmit} 
\end{remark}

With Proposition~\ref{prop:optimal power allocation for ESR OMA-II}, given a triple \((\lambda,\delta,\mu)\), \(\mathrm{(P1\text{-}OMA\text{-}II\text{-}sub)}\) is first solved state by state; then by updating \((\lambda,\delta,\mu)\) in accordance with the associated sub-gradient \((\bar P-\mb{E}_\nu[p_k(\nu)+p_{\bar k}(\nu)], \mb{E}_\nu[R_k^{\rm OMA\text{-}II}(\nu)]-\bar R, \mb{E}_\nu[R_{\bar k}^{\rm OMA\text{-}II}(\nu)]-\bar R)^T\), \(\mathrm{(P1\text{-}OMA\text{-}II)}\) is iteratively solved.  

Next, we rigorously prove that the ESR achieved by OMA-Type-II cannot perform better than that achieved by NOMA. To prove so, we denote the optimal power and time/frequency allocation to \(\mathrm{(P1\text{-}OMA\text{-}II)}\) by \(\{p_1^\ast(\nu),p_2^\ast(\nu),\alpha_1^\ast(\nu)\}\). Then let \(p_1=p_1^\ast\) and \(p_2=p_2^\ast\) in each fading state for NOMA transmission. If an alternative user is selected to transmit, then the optimum \(\alpha_k\) associated with it is seen to be \(1\) (c.f.~\eqref{eq:optimal power allocation for ESR OMA-II}). Hence, assuming \(p_1^\ast>0\) (\(\alpha_1^\ast=1\)) and \(p_2^\ast=0\) (\(\alpha_2^\ast=0\)), \(R_1^{\rm NOMA}\) turns out be \(\log_2(1+p_1^\ast g_1)\) and \(R_2^{\rm NOMA}=0\), which is exactly equal to \(R_1^{\ast\rm OMA\text{-}II}\) and \(R_2^{\ast\rm OMA\text{-}II}\), respectively; vice versa when \(p_1^\ast=0\) (\(\alpha_1^\ast=0\)) and \(p_2^\ast>0\) (\(\alpha_2^\ast=1\)). The other trivial case is that \(R_k^{\rm NOMA}=R_k^{\ast\rm OMA\text{-}II}=0\), \(\forall k\), when \(p_1^\ast=p_2^\ast=0\). Hence, with \(p_1=p_1^\ast\) and \(p_2=p_2^\ast\) in each fading state, \(\mb E_\nu[R_k^{\rm NOMA}(\nu)]=\mb E_\nu[R_k^{\ast\rm OMA\text{-}II}(\nu)]\ge\bar R\), \(\forall k\), is met. It is also easily examined that \eqref{eq:min rate constraint under full CSIT} is satisfied. To sum up, the optimal solution to \(\mathrm{(P1\text{-}OMA\text{-}II)}\) proves to be feasible to \(\mathrm{(P1\text{-}NOMA)}\), the former of which thus yields an optimum value no more than the latter.

\subsection{Partial CSIT}
By analogy, the partial CSIT counterpart of Problem \(\mathrm{(P1\text{-}XX)}\) is formulated as below:
\begin{subequations} 
\begin{align}\mathrm{(P1^\prime\text{-}XX)}:~\mathop{\mathtt{Maximize}}_{p_s,p_w,\alpha_k}
& ~~~ \mb{E}_\nu[R_k^{\prime\rm XX}(\nu)+R_{\bar k}^{\prime\rm XX}(\nu)]\notag\\
\mathtt {Subject \ to}& ~~~p_s+p_w\leq\bar P,\label{eq:APC under partial CSIT}\\
& ~~~p_s\ge 0, \ p_w\ge 0, \\
 &~~~\mb{E}_\nu[R_k^{\prime\rm XX}(\nu)]\ge\bar R^\prime, \; \forall k,\label{eq:min rate constraint under partial CSIT}
\end{align}   
\end{subequations} where \(\alpha_k\)'s is only valid in the transmission adopting OMA-Type-II. Similar to Problem \(\mathrm{(P1\text{-}XX)}\), \eqref{eq:min rate constraint under partial CSIT} constrain the minimum average user rate achieved by the two users. In the following, we provide optimal solution to  \(\mathrm{(P1^\prime\text{-}NOMA)}\) and \(\mathrm{(P1^\prime\text{-}OMA\text{-}II)}\), respectively.
\subsubsection{Optimal Solution to \(\mathrm{(P1^\prime\text{-}NOMA)}\)}\label{subsubsec:Optimal Solution to (P1'-NOMA) under Partial CSIT}
Since only the relation between the two users' channel gains at each fading state and their CDI are known to the BS, we first derive the expectation of \(R_k^{\prime\rm NOMA}(\nu)\)'s as function of \(p_s\) and \(p_w\), and then solve \(\mathrm{(P1^\prime\text{-}NOMA)}\) in accordance with these expectation results. 

First, denote the RV \(|h_k(\nu)|^2\) (\(|h_{\bar k}(\nu)|^2\)) by \(X\) (\(Y\))\footnote{Note that we assume \(\sigma_k^2=\sigma^2\), \(\forall k\), throughout the paper such that the relation between the effective channels of the two users , i.e., \(g_k(\nu)\) and \(g_{\bar k}(\nu)\), is equivalent to that between \(|h_k(\nu)|^2\) and \(|h_{\bar k}(\nu)|^2\).}. Also, denote the SNR \(p_sg_k(\nu)\) and the SINR \(\Myfrac{p_wg_k(\nu)}{(p_sg_k(\nu)+1)}\) (c.f.~\eqref{eq:R_k prime NOMA under partial CSIT}) by \(\Gamma_k\) and \(\tilde{\Gamma}_k\), respectively. It thus follows that the conditional cumulative density functions (CDFs) of \(\Gamma_k\) and \(\tilde{\Gamma}_k\) are given by\textcolor{black}{\footnote{These results are not explicitly applicable to $K>2$ case. A better approach to deal with the more general cases with $K>2$  is to solicit order statistics \cite{ding2014performance,Liu2016TVT}, which is beyond the scope of this treatise.}}
\begin{align}
{F_{{\Gamma _{k|X \ge Y}}}}\left( {{z}} \right) =& \frac{\Pr\{\Myfrac{p_sX}{\sigma_k^2}\le z,X \ge Y\}}{\Pr\{X \ge Y\}}\notag\\
= & 1 - \frac{\lambda_k+\lambda_{\bar k}}{\lambda_{\bar k}}{e^{ - {\lambda _k}{\varepsilon_k}}} + \frac{\lambda _k}{\lambda_{\bar k}}e^{-\left(\lambda_k+\lambda_{\bar k}\right)\varepsilon_k}, \label{eq:conditional CDF of Gamma_k in NOMA}\\
{F_{{\tilde\Gamma_k|X<Y }}}\left( {{z}} \right) = & \frac{\Pr\{\Myfrac{p_wX}{\left(p_sX+\sigma_k^2\right)}\le z,X < Y\}}{\Pr\{X < Y\}}\notag\\
=& \left\{ \begin{array}{l}
1, \ \ \mbox{if}\ {p_w} - {p_s}{z} \le 0,\\
1 - {e^{ - \left( {{\lambda _k} + {\lambda _{\bar k}}} \right){\tilde\varepsilon_k}}}, \ \  \mbox{otherwise,}\end{array}, \right. \label{eq:conditional CDF of tilde Gamma_k in NOMA}
\end{align} respectively, where \(\varepsilon_k\triangleq\frac{\sigma_k^2z}{p_s}\) and \(\tilde\varepsilon_k\triangleq\frac{\sigma_k^2z}{p_w-p_sz}\).
In accordance with \eqref{eq:conditional CDF of Gamma_k in NOMA} and \eqref{eq:conditional CDF of tilde Gamma_k in NOMA}, \(\mb{E}_\nu[R_k^{\prime\rm NOMA}(\nu)]\)'s can be obtained by the following proposition.
\begin{proposition}
The ergodic rate for NOMA user \(\mathcal{U}_k\), \(k\in\{1,2\}\), under partial CSIT is given by
\begin{multline}
\mb{E}_\nu[R_k^{\prime\rm NOMA}(\nu)]=\frac{2}{{\ln 2}}\frac{{{\lambda _k}}}{{{\lambda _k} + {\lambda _{\bar k}}}}f\left(\frac{{\left( {{\lambda _k} + {\lambda _{\bar k}}} \right)\sigma _k^2}}{{{p_s}}} \right )\\
- \frac{1}{{\ln 2}}\frac{{{\lambda _k}}}{{{\lambda _k} + {\lambda _{\bar k}}}}f\left(\frac{{\left( {{\lambda _k} + {\lambda _{\bar k}}} \right)\sigma _k^2}}{{{p_s+p_w}}} \right )
- \frac{1}{{\ln 2}}f\left(\frac{{{\lambda _k}\sigma _k^2}}{{{p_s}}} \right ), \label{eq:ER for the kth user in NOMA}
\end{multline} where \(f(\cdot)\) denotes the function \(f(x)=e^x{\rm Ei}(-x)\) (\(x>0\)).
\label{prop:ER for the kth user in NOMA}
\end{proposition}
\begin{IEEEproof}
Please refer to Appendix~\ref{appendix:proof of ER for the kth user in NOMA}.
\end{IEEEproof}
Since the optimization variable \(p_w\) only contributes to \(f(\frac{{\left( {{\lambda _k} + {\lambda _{\bar k}}} \right)\sigma _k^2}}{{{p_s} + {p_w}}})\), we examine the property of \(\mb{E}_\nu[{R_k^{\prime\rm NOMA}(\nu)}]\) in terms of \(p_w\) by studying \(f(\frac{{\left( {{\lambda _k} + {\lambda _{\bar k}}} \right)\sigma _k^2}}{{{p_s} + {p_w}}})\) as follows:
\begin{align}
\frac{{\partial f\left(\frac{{\left( {{\lambda _k} + {\lambda _{\bar k}}} \right)\sigma _k^2}}{{{p_s} + {p_w}}}\right)}}{{\partial {p_w}}} = & \frac{{ \partial \left( {{e^{\frac{{\left( {{\lambda _k} + {\lambda _{_{\bar k}}}} \right)\sigma _k^2}}{{{p_s} + {p_w}}}}}{\rm Ei}\left( { - \frac{{\left( {{\lambda _k} + {\lambda _{_{\bar k}}}} \right)\sigma _k^2}}{{{p_s} + {p_w}}}} \right)} \right)}}{{\partial {p_w}}}\notag\\
=  & {\frac{1}{{{p_s} + {p_w}}}\left( { u{\rm E1}\left( u \right){e^u} - 1} \right)}\notag\\
\stackrel{(a)}{<} & {\frac{1}{{{p_s} + {p_w}}}\left(u\ln\left(1+\frac{1}{u}\right)-1\right)}< 0, \label{eq:partial derivitive of A1 w.r.t. pw}
\end{align} 
where \(\frac{{({\lambda _k} + {\lambda _{\bar k}})\sigma _k^2}}{{{p_s} + {p_w}}}\triangleq u\), \({\rm E1}(x)=-{\rm Ei}(-x)\) (\(x>0\)), and \((a)\) is due to the inequality \({\rm E1}(x)e^x<\ln(1+\frac{1}{x})\) (\(X>0\)) \cite[Eq.~(5.1.20)]{abramowitz1972handbook}. Hence, \(\mb{E}_\nu[{R_k^{\prime\rm NOMA}(\nu)}]\)'s proves to monotonically increase with \(p_w\). 

Next, we solicit this monotonicity for solving \(\mathrm{(P1^\prime\text{-}NOMA)}\). As it is easily seen that given any \(p_s\), \(\mathrm{(P1^\prime\text{-}NOMA)}\)  attains its optimum value when \(\mb{E}_\nu[{R_k^{\prime\rm NOMA}(\nu)}]\)'s takes on its maximum w.r.t. \(p_w\), i.e., when \(p_w=\bar P-p_s\) (c.f.~\eqref{eq:APC under partial CSIT}), \(\mathrm{(P1^\prime\text{-}NOMA)}\) is thus related with only one optimization variable \(p_s\). Hence, one-dimension search over \(p_s\in[0,\bar P]\) can be implemented to find the optimum solution (up to numerical accuracy) to \(\mathrm{(P1^\prime\text{-}NOMA)}\). 
%In addition, the feasibility of \(\mathrm{(P1^\prime\text{-}NOMA)}\) can be similarly studied as that of \(\mathrm{(P1\text{-}NOMA)}\).
\subsubsection{Optimal Solution to \(\mathrm{(P1^\prime\text{-}OMA\text{-}II)}\)}\label{subsubsec:Optimal Solution to (P1'-OMA-II) under Partial CSIT}
Denoting the SNR of \(\mathcal{U}_k\) in the case of \(X\ge Y\) by \(\Gamma_k\), and that in the case of \(X<Y\) by \(\tilde\Gamma_k\) (c.f.~\eqref{eq:R_k prime OMA-II under partial CSIT}), the conditional CDFs of \(\Gamma_k\) and \(\tilde\Gamma_k\) are given by 
\begin{align}
{F_{{\Gamma _{k|X \ge Y}}}}\left( {{z}} \right)= & 1 - \frac{\lambda_k+\lambda_{\bar k}}{\lambda_{\bar k}}{e^{ - {\lambda _k}{\varphi_k}}} + \frac{\lambda _k}{\lambda_{\bar k}}e^{-\left(\lambda_k+\lambda_{\bar k}\right)\varphi_k},\label{eq:conditional CDF of Gamma_k in OMA-II}\\
{F_{{\tilde\Gamma _{k|X < Y}}}}\left( {{z}} \right)= & 1-e^{-\left(\lambda_k+\lambda_{\bar k}\right)\tilde\varphi_k}, \label{eq:conditional CDF of tilde Gamma_k in OMA-II}
\end{align} where \(\varphi_k\triangleq\frac{\alpha_k\sigma_k^2z}{p_s}\) and \(\tilde\varphi_k\triangleq\frac{\alpha_k\sigma_k^2z}{p_w}\). With \eqref{eq:conditional CDF of Gamma_k in OMA-II} and \eqref{eq:conditional CDF of tilde Gamma_k in OMA-II}, we have the following proposition.
\begin{proposition}
The ergodic rate for user \(\mathcal{U}_k\) operating with OMA-type-II, \(k\in\{1,2\}\), under partial CSIT is given by
\begin{multline}
\mb{E}_\nu[R_k^{\prime\rm OMA\text{-}II}(\nu)]=\\
\frac{\alpha_k}{{\ln 2}}\left(-f\left(\frac{{ {{\lambda _k}}\alpha_k\sigma _k^2}}{{{p_s}}} \right )+\frac{\lambda_k}{\lambda_k+\lambda_{\bar k}}f\left(\frac{{ {{\left(\lambda _k+\lambda _{\bar k} \right )}}\alpha_k\sigma _k^2}}{{{p_s}}} \right )\right.\\
\left.-\frac{\lambda_k}{\lambda_k+\lambda_{\bar k}}f\left(\frac{{ {{\left(\lambda _k+\lambda _{\bar k} \right )}}\alpha_k\sigma _k^2}}{{{p_w}}} \right )\right ). \label{eq:ER for the kth user in OMA-II} 
\end{multline} \label{prop:ER for the kth user in OMA-II} 
\end{proposition}
\begin{IEEEproof}
Please refer to Appendix~\ref{appendix:proof of ER for the kth user in OMA-II}.
\end{IEEEproof}

Similar as is done in \eqref{eq:partial derivitive of A1 w.r.t. pw}, \(\mb{E}_\nu[R_k^{\prime\rm OMA\text{-}II}(\nu)]\)'s can be shown to monotonically increase with \(p_w\) as well. Therefore it implies that the optimal solution to \(\mathrm{(P1^\prime\text{-}OMA\text{-}II)}\) satisfies \(p_s+p_w=\bar P\). As a result, there are two optimization variables (\(p_s\) and \(\alpha_k\)) remaining for \(\mathrm{(P1^\prime\text{-}OMA\text{-}II)}\), which can be solved (up to numerical accuracy) by two-dimension search over \(\{(p_s,\alpha_k)\vert p_s\in[0,\bar P],\alpha_k\in[0,1]\}\) such that \(\mb{E}_\nu[R_k^{\prime\rm OMA\text{-}II}(\nu)]\ge\bar R^\prime\), \(k\in\{1,2\}\).

\section{Optimum Delay-Limited Transmission}\label{sec:Optimum Delay-Limited Transmission}
In delay-limited scenarios, each user attempts to maintain their respective prescribed rate in as much fading states as possible so as to reduce their outage probability (c.f.~\eqref{eq:zeta_k NOMA for strong user under full CSIT}, \eqref{eq:zeta_k NOMA for weak user under full CSIT}, and \eqref{eq:sigma_k OMA-II under full CSIT}). When the users compete for power and/or time/frequency resources to get their intended data transmitted at the target rate in each fading state, the combined effects of outage probability and individual target rate accounts for the DLT of each user, which causes the solution to the sum of DLT maximization non-trivial. In this section, the optimal trade-offs between the system sum-DLT and the maximum outage probability requirement for the users is investigated for different multiple access schemes under full and partial CSIT, respectively. \textcolor{black}{Particularly, under full CSIT, the DLT maximization problems are solved using Lagrangian dual decomposition levering ``time-sharing'' conditions, while under partial CSIT, the individual user's outage probability needs to be first analysed in closed form by means of CDI.}

\subsection{Full CSIT}
In the case of full CSIT, we aim for maximizing the system sum of DLT by jointly optimizing the individual transmit power as well as time/frequency allocation over different fading states, subject to a given pair of APC and PPC at the BS, and a maximum user outage probability constraint. The optimization problem is thus formulated as below.
\begin{subequations}
\begin{align}
\mathrm{(P2\text{-}XX)}:&~~\mathop{\mathtt{Maximize}}_{\{p_k(\nu), p_{\bar k}(\nu),\alpha_k(\nu)\}}\bar R_k(1-\zeta_k^{\rm XX})+\bar R_{\bar k}(1-\zeta_{\bar k}^{\rm XX})\notag\\
&~~\mathtt{Subject \ to}\notag\\
&~~\mb{E}_\nu[p_k(\nu)+p_{\bar k}(\nu)]\leq \bar P,\\
&~~p_k(\nu)+p_{\bar k}(\nu)\leq \hat P, \ \forall\nu,\\
&~~p_k(\nu)\ge 0, \ p_{\bar k}(\nu)\ge 0,\ \forall\nu,\\
&~~\mb{E}_\nu[X_k^{\rm XX}(\nu)]\le\bar\zeta, \; \forall k,\label{eq:max outage constraint under full CSIT}  
\end{align}
\end{subequations} where \(\{\alpha_k(\nu)\}\)'s are only valid when the two users access the channel by OMA-Type-II. It is worthy of noting that given the same target rate intended for each user, i.e., \(\bar R_k=\bar R_{\bar k}=\bar R\), even if \(\mathcal{U}_k\) and \(\mathcal{U}_{\bar k}\) suffer from ``near-far'' physical condition, the far user can still successfully decode its data at this constant rate for more than  \(1-\bar\zeta\) proportion of the fading states, thanks to the constraints \eqref{eq:max outage constraint under full CSIT}. As seen from \eqref{eq:X_k NOMA for strong user under full CSIT} and \eqref{eq:X_k NOMA for weak user under full CSIT} (\eqref{eq:X_k OMA-II under full CSIT}), the discrete value of \(X_k^{\rm NOMA}(\nu)\) (\( X_k^{\rm OMA\text{-}II}(\nu)\))'s renders non-convexity w.r.t. the optimization variables \(p_k(\nu)\), \(k\in\{1,2\}\), and thus Problem \(\mathrm{(P2\text{-}XX)}\) is also non-convex. Therefore we exploit the similar ``time-sharing'' condition aforementioned to find their optimal solutions in subsection~\ref{subsubsec:Optimal Solution to (P2-NOMA) under Full CSIT} and \ref{subsubsec:Optimal Solution to (P2-OMA-II) under Full CSIT}, respectively.  In the following, we aim for solving \(\mathrm{(P2\text{-}NOMA)}\) and \(\mathrm{(P2\text{-}OMA\text{-}II)}\), respectively.

\subsubsection{Optimal Solution to \(\mathrm{(P2\text{-}NOMA)}\)}\label{subsubsec:Optimal Solution to (P2-NOMA) under Full CSIT}
Adopting Lagrangian dual decomposition method, the Lagrangian of Problem \(\mathrm{(P2\text{-}NOMA)}\) is given by
\begin{multline}
\mathcal{L}_2^{\rm NOMA}(\{p_k(\nu)\},\{p_{\bar k}(\nu)\},\lambda,\delta,\mu)=\\
\mb{E}_\nu[-\bar R_kX_k^{\rm NOMA}(\nu)-\bar R_{\bar k}X_{\bar k}^{\rm NOMA}(\nu)-\lambda(p_k(\nu)+p_{\bar k}(\nu))-\\
\delta X_k^{\rm NOMA}(\nu)-\mu X_{\bar k}^{\rm NOMA}(\nu)]
+\lambda\bar P+\delta\bar\zeta+\mu\bar\zeta, \label{eq:Lagrangian of (P2-NOMA)}
\end{multline} where \(\lambda\) is the Lagrangian multiplier associated with the APC; \(\delta\) and \(\mu\) are those associated with the maximum user outage probability constraints given in \eqref{eq:max outage constraint under full CSIT} for \(\mathcal{U}_k\) and \(\mathcal{U}_{\bar k}\), respectively. In line with the principle of dual decomposition, \eqref{eq:Lagrangian of (P2-NOMA)} can be maximized by decoupling it into independent subproblems each for one fading state and solving those subproblems in parallel. Define \(\bar{\mathcal{L}}_2^{\rm NOMA}(p_k,p_{\bar k})=\bar R_kX_k^{\rm NOMA}+\bar R_{\bar k}X_{\bar k}^{\rm NOMA}+\lambda(p_k+p_{\bar k})+\delta X_k^{\rm NOMA}+\mu X_{\bar k}^{\rm NOMA}\). With the fading index \(\nu\) safely dropped, given the dual variables' triple \((\lambda,\delta,\mu)\), the following problem is typical of the subproblems sharing the same structure:
\begin{align*}
\mathrm{(P2\text{-}NOMA\text{-}sub)}:~\mathop{\mathtt{Minimize}}_{p_k\ge 0,p_{\bar k}\ge 0}
& ~~~\bar{\mathcal{L}}_2^{\rm NOMA}(p_k,p_{\bar k})\\
\mathtt {Subject \ to}& ~~~p_k+p_{\bar k}\leq\hat P.
\end{align*}

Then, we investigate the possible combinations of outage occurrences for  \(\mathcal{U}_k\) and \(\mathcal{U}_{\bar k}\). Assuming \(g_k>g_{\bar k}\), \(k\in\{1,2\}\), the possible combinations of indicator function \(X_k^{\rm NOMA}\) and the corresponding decoding strategies adopted by \(\mathcal{U}_k\) are summarized in Table~\ref{table:combinations of U1 and U2's NOMA decoding strategies}, where \(\mathcal{U}_{\bar k}\rightarrow\mathcal{U}_{\bar k}\) represents that \(\mathcal{U}_{\bar k}\) directly decodes its own information TIAN; \(\mathcal{U}_k\rightarrow\mathcal{U}_{\bar k}\rightarrow\mathcal{U}_k\) denotes \(\mathcal{U}_k\)'s attempt to perform SIC\footnote{Given \(g_k>g_{\bar k}\), we assume the decoding order \(\mathcal{U}_k\rightarrow\mathcal{U}_{\bar k}\rightarrow\mathcal{U}_k\), since in delay-limited NOMA, it is optimum to have only the user with better CSI perform SIC for the sake of saving the total transmit power.}; \(\not\rightarrow\) indicates failure of decoding. Specifically, if the first step succeeds, \(\mathcal{U}_k\) is able to cancel the interference from \(\mathcal{U}_{\bar k}\), otherwise \(\mathcal{U}_k\) continues to decode its own treating  \(\mathcal{U}_{\bar k}\)'s as interference. Based on Table~\ref{table:combinations of U1 and U2's NOMA decoding strategies}, we derive the optimal solution to \(\mathrm{(P2\text{-}NOMA\text{-}sub)}\) in the following proposition.   
{\footnotesize% Please add the following required packages to your document preamble:
\begin{table}[htp]
\centering
%\caption{Possible combinations of two-user decoding strategies for delay-limited transmission assuming \(g_1>g_2\).}
\caption{}\label{table:combinations of U1 and U2's NOMA decoding strategies}
\resizebox{.45\textwidth}{!}{%
\begin{tabular}{@{}lll@{}}
\toprule
\(\mathcal{U}_k\) & \(\begin{aligned}[t]\mathcal{U}_k\not\rightarrow \mathcal{U}_{\bar k}\not\rightarrow \mathcal{U}_k,\, X_k^{\rm NOMA}=1\ \ ({\rm I.A})\\ \mathcal{U}_k\not\rightarrow \mathcal{U}_{\bar k}\rightarrow \mathcal{U}_k,\, X_k^{\rm NOMA}=0\ \ ({\rm I.B}) \\ \mathcal{U}_k\rightarrow \mathcal{U}_{\bar k}\not\rightarrow \mathcal{U}_k,\, X_k^{\rm NOMA}=1\ \ ({\rm I.C})\\ \mathcal{U}_k\rightarrow \mathcal{U}_{\bar k}\rightarrow \mathcal{U}_k,\, X_k^{\rm NOMA}=0\ \ ({\rm I.D})
\end{aligned}\) & \(\begin{aligned}[t]\mathcal{U}_k\rightarrow \mathcal{U}_{\bar k}\not\rightarrow \mathcal{U}_k,\, X_k^{\rm NOMA}=1\ \ ({\rm II.A}) \\  \mathcal{U}_k\rightarrow \mathcal{U}_{\bar k}\rightarrow \mathcal{U}_k,\, X_k^{\rm NOMA}=0\ \ ({\rm II.B}) 
\end{aligned}\) \\ \midrule
\(\mathcal{U}_{\bar k}\) & \(\mathcal{U}_{\bar k}\not\rightarrow \mathcal{U}_{\bar k},\ X_{\bar k}^{\rm NOMA}=1\) & \(\mathcal{U}_{\bar k}\rightarrow \mathcal{U}_{\bar k},\ X_{\bar k}^{\rm NOMA}=0\)\\ \bottomrule
\end{tabular}%
}
\end{table}}

\begin{proposition}
The optimal power allocation to Problem \(\mathrm{(P2\text{-}NOMA\text{-}sub)}\) assuming \(g_k>g_{\bar k}\) is given by
\begin{align*}
(p_k^\ast, p_{\bar k}^\ast)=\arg\min\limits_{i\in\{1,2,3,4\}}\{\bar{\mathcal{L}}_2^{\rm NOMA}(p_{i,k},p_{i,\bar k})\mathbbm{1}_{p_{i,k}+p_{i,\bar k}}\},
\end{align*} where \(p_{i,k}\)'s and \(p_{i,\bar k}\)'s are given by\textcolor{black}{\footnote{In fact, under full CSIT, following the same method utilized to develop Proposition~\ref{prop:optimal power allocation for DLT NOMA}, the two-user results can be generalized to cases with $K>2$ by mathematical induction.}} 
\begin{align}
\kern-4pt\left\{\begin{array}{l} 
p_{1,k}=0, \ p_{1,\bar k}=0\\ 
p_{2,k}=\frac{2^{\bar R_k}-1}{g_k}, \  p_{2,\bar k}=0\\
p_{3,k}=0, \ p_{3,\bar k}=\frac{2^{\bar R_{\bar k}}-1}{g_{\bar k}}\\
p_{4,k}=\frac{2^{\bar R_k}-1}{g_k}, \ p_{4,\bar k}=\left(2^{\bar R_{\bar k}}-1 \right )\left(\frac{2^{\bar R_k}-1}{g_k}+\frac{1}{g_{\bar k}}\right ) \end{array},\right.\label{eq:case-specified optimal power allocation for DLT NOMA under full CSIT}
\end{align} and the indicator function \(\mathbbm{1}_{(\cdot)}\) is defined the same as \eqref{eq:power constraint indicator}.
\label{prop:optimal power allocation for DLT NOMA}
\end{proposition}
\begin{IEEEproof}
To minimize \(\bar{\mathcal{L}}_2^{\rm NOMA}(p_k,p_{\bar k})\), we need to examine every case of combination regarding \(\mathcal{U}_k\)'s and \(\mathcal{U}_{\bar k}\)'s outage occurrences so as to find the one that minimizes \(\bar{\mathcal{L}}_2^{\rm NOMA}(p_k,p_{\bar k})\). First, we show that the cases I.C and I.D can be safely removed since they are always outperformed by other cases. Take I.C as an example, if \(\mathcal{U}_k\) succeeds in decoding \(\mathcal{U}_{\bar k}\)'s message at rate \(\bar R_{\bar k}\), it inexplicitly suggests that \(\mathcal{U}_{\bar k}\)'s message is transmitted at \(p_{\bar k}>0\). Therefore, the corresponding \(\bar{\mathcal{L}}_2^{\rm NOMA}(p_k,p_{\bar k})=\bar R_k+\bar R_{\bar k}+\lambda p_{\bar k}+\delta+\mu\) is strictly larger than \(\bar{\mathcal{L}}_2^{\rm NOMA}(0,p_{\bar k})=\bar R_k+\bar R_{\bar k}+\delta+\mu\) in Case I.A. Similarly, Case I.D can be shown to be strictly  outperformed by Case I.B. With the remaining four cases, \(p_{i,k}\) (\(p_{i,\bar k}\)), \(i\in\{1,2,3,4\}\), is the minimum power required for \(\mathcal{U}_k\) (\(\mathcal{U}_{\bar k}\)) to succeed in transmission associated with the case I.A, I.B, II.A, and II.B, respectively. Next, select the minimizer out from these four cases, which depends on how the required transmit power weighs \(\bar R_k\)'s as well as the given multipliers \((\lambda,\delta,\mu)\).
\end{IEEEproof}

Note from Proposition~\ref{prop:optimal power allocation for DLT NOMA} that the optimal power policy allocates either the minimum required power to support \(\mathcal{U}_{k}\) and/or \(\mathcal{U}_{\bar k}\)'s transmission at their respective target rate or completely shuts down the transmission. For example, when \(\mathcal{U}_k\) suspends its transmission in Cases I.A and II.A, Case II.A outperforms Case I.A if and only if (iff) \(p_{\bar k}\le\frac{\bar R_{\bar k}+\mu}{\lambda}\). From the perspective of fairness, when \(\mu\) is large enough appealing for smaller outage, this condition is easier to be satisfied and thus \(X_{\bar k}^{\rm NOMA}\) is more likely to be \(0\), and vice versa.    

With Proposition~\ref{prop:optimal power allocation for DLT NOMA}, given any multiplier-triple \((\lambda,\delta,\mu)\), the maximum of the Lagrangian in \eqref{eq:Lagrangian of (P2-NOMA)} is obtained by solving \(\mathrm{(P2\text{-}NOMA\text{-}sub)}\) state by state in parallel. Finally, \(\mathrm{(P2\text{-}NOMA)}\) is solved by updating \((\lambda,\delta,\mu)\) in accordance with the ellipsoid method.

\subsubsection{Optimal Solution to \(\mathrm{(P2\text{-}OMA\text{-}II)}\)}\label{subsubsec:Optimal Solution to (P2-OMA-II) under Full CSIT}
Despite of its non-convexity  due to the same reason as that for \(\mathrm{(P2\text{-}NOMA)}\), we can still find the optimal solution to \(\mathrm{(P2\text{-}OMA\text{-}II)}\) thanks to the ``time-sharing'' condition that \(\mathrm{(P2\text{-}OMA\text{-}II)}\) meets. 

Similar to Section~\ref{subsubsec:Optimal Solution to (P2-NOMA) under Full CSIT}, \((\mathrm{(P2\text{-}OMA\text{-}II)}\) can also be decoupled into as many subproblems as the number of fading states each for one fading state, which is expressed as
\begin{align*}
\mathrm{(P2\text{-}OMA\text{-}II\text{-}sub)}:&~~\mathop{\mathtt{Minimize}}_{p_k\ge 0,p_{\bar k}\ge 0,\alpha_k}\bar{\mathcal{L}}_2^{\rm OMA\text{-}II}(p_k,p_{\bar k},\alpha_k)\notag\\
&~~\mathtt {Subject \ to}\\
&~~p_k+p_{\bar k}\leq\hat P,\\
&~~0\le\alpha_k\le 1, \; \forall k,
\end{align*}    
where the objective function is defined as \(\bar{\mathcal{L}}_2^{\rm OMA\text{-}II}(p_k,p_{\bar k},\alpha_k)=\bar R_k X_k^{\rm OMA\text{-}II}+\bar R_{\bar k}X_{\bar k}^{\rm OMA\text{-}II}+\lambda(p_k+p_{\bar k})+\delta X_k^{\rm OMA\text{-}II}+\mu X_{\bar k}^{\rm OMA\text{-}II}\) with the fading index \(\nu\) dropped for brevity.

Since each \(\mathcal{U}_{k}\) only needs to decode its own information without seeing interference in the orthogonal transmission, the possible combinations of outage occurrences for \(\mathcal{U}_{k}\) and \(\mathcal{U}_{\bar k}\) are easily shown in Table~\ref{table:combinations of U1 and U2's OMA-II decoding strategies}, where \(\mathcal{U}_{k}\rightarrow\mathcal{U}_{k}\) denotes \(\mathcal{U}_{k}\)'s direct decoding of its own message, \(k\in\{1,2\}\).   
{\footnotesize% Please add the following required packages to your document preamble:
\begin{table}[htp]
\centering
%\caption{Possible combinations of two-user decoding strategies for delay-limited transmission assuming \(g_1>g_2\).}
\caption{}\label{table:combinations of U1 and U2's OMA-II decoding strategies}
\resizebox{.45\textwidth}{!}{%
\begin{tabular}{@{}lll@{}}
\toprule
\(\mathcal{U}_k\) & \(\begin{aligned}[t]\mathcal{U}_k\not\rightarrow\mathcal{U}_k,\, X_k^{\rm OMA\text{-}II}=1\\ \mathcal{U}_k\rightarrow\mathcal{U}_k,\, X_k^{\rm OMA\text{-}II}=0
\end{aligned}\) & \(\begin{aligned}[t]\mathcal{U}_k\not\rightarrow\mathcal{U}_k,\, X_k^{\rm OMA\text{-}II}=1\\  \mathcal{U}_k\rightarrow\mathcal{U}_k,\, X_k^{\rm OMA\text{-}II}=0
\end{aligned}\) \\ \midrule
\(\mathcal{U}_{\bar k}\) & \(\mathcal{U}_{\bar k}\not\rightarrow\mathcal{U}_{\bar k},\ X_{\bar k}^{\rm OMA\text{-}II}=1\) & \(\mathcal{U}_{\bar k}\rightarrow\mathcal{U}_{\bar k},\ X_{\bar k}^{\rm OMA\text{-}II}=0\)\\ \bottomrule
\end{tabular}%
}
\end{table}}
Based on Table~\ref{table:combinations of U1 and U2's OMA-II decoding strategies}, we obtain the optimal solution to \(\mathrm{(P2\text{-}OMA\text{-}II\text{-}sub)}\) in the following proposition.

\begin{proposition}
The optimal power allocation to Problem \(\mathrm{(P2\text{-}OMA\text{-}II\text{-}sub)}\) is given by
\begin{multline}
(p_k^\ast, p_{\bar k}^\ast,\alpha_k^\ast)=\\
\arg\min\limits_{i\in\{1,2,3,4\}}\{\bar{\mathcal{L}}_2^{\rm OMA\text{-}II}(p_{i,k},p_{i,\bar k},\alpha_{i,k})\mathbbm{1}_{p_{i,k}+p_{i,\bar k}}\},
\end{multline} where \(p_{i,k}\)'s and \(p_{i,\bar k}\)'s are given by
\begin{align}
\kern-4pt\left\{\begin{array}{l} 
\kern-4pt p_{1,k}=0, \ p_{1,\bar k}=0, \ \alpha_{1,k}=0;\\ 
\kern-4pt p_{2,k}=\frac{2^{\bar R_k}-1}{g_k}, \ p_{2,\bar k}=0, \ \alpha_{2,k}=1;\\
\kern-4pt p_{3,k}=0, \ p_{3,\bar k}=\frac{2^{\bar R_{\bar k}}-1}{g_{\bar k}}, \ \alpha_{3,k}=0;\\
\kern-4pt p_{4,k}=\frac{\alpha^\ast_k(2^{\frac{\bar R_k}{\alpha^\ast_k}}-1)}{g_k}, \ p_{4,\bar k}=\frac{\alpha_{\bar k}^\ast(2^{\frac{\bar R_{\bar k}}{\alpha_{\bar k}^\ast}}-1)}{g_{\bar k}} , \ \alpha_{4,k}=\alpha^\ast_k. \end{array}\right.\label{eq:case-specified optimal power allocation for DLT OMA-II under full CSIT}
\end{align} In \eqref{eq:case-specified optimal power allocation for DLT OMA-II under full CSIT}, \(\alpha_k^\ast\)'s denotes the optimum proportion of time/frequency resource allocated to \(\mathcal{U}_k\)'s to minimize the instantaneous total transmit power, which is obtained by solving the following (convex) problem.
\begin{align*}
&\mathrm{(P2\text{-}OMA\text{-}II\text{-}MP)}:\notag\\
&\mathop{\mathtt{Minimize}}_{\alpha_k}~~
\frac{\alpha_k(2^{\frac{\bar R_k}{\alpha_k}}-1)}{g_k}+\frac{\alpha_{\bar k}(2^{\frac{\bar R_{\bar k}}{\alpha_{\bar k}}}-1)}{g_{\bar k}}\\
&\mathtt {Subject \ to}~~0\le\alpha_k\le 1.
\end{align*}
In addition, the indicator function is given by \eqref{eq:power constraint indicator}.\label{prop:optimal power allocation for DLT OMA-II}
\end{proposition}
\begin{IEEEproof}
Please refer to Appendix~\ref{appendix:proof of optimal power allocation for DLT OMA-II}. 
\end{IEEEproof}

As a result, given any triple \((\lambda,\delta,\mu)\), \(\mathrm{(P2\text{-}OMA\text{-}II\text{-}sub)}\) is solved. \(\mathrm{(P2\text{-}OMA\text{-}II)}\) is then solved by updating \((\lambda,\delta,\mu)\) using the ellipsoid method similarly as solving \(\mathrm{(P2\text{-}NOMA)}\).

Next, given the same set of \(\bar R_k\)'s, we provide mathematical proof for the superiority of NOMA over OMA-Type-II in terms of the optimum sum of DLT. To prove so, we introduce the following lemma.
\begin{lemma}
Given \(g_1\ge g_2\ge\ldots\ge g_K>0\), denoting \(\min\limits_{\sum_{i=1}^K\alpha_i=1}\sum_{i=1}^K\frac{(2^{\frac{\bar R_i}{\alpha_i}}-1)\alpha_i}{g_i}\) by \(P_{\rm O2}^\ast\), and \(\sum_{i=0}^{K-1}\frac{(2^{\bar R_{K-i}}-1)2^{\sum_{j=0}^{i-1}\bar R_{K-j}}}{g_{K-i}}\) by \(P_{\rm N}^\ast\), where \(\alpha_i\ge 0\) and \(\bar R_i\ge 0\), \(\forall i\), then it follows that \(P_{\rm O2}^\ast\ge P_{\rm N}^\ast\). \label{lemma:power saving for DLT NOMA over OMA-II}
\end{lemma} 
\begin{IEEEproof}
Please refer to Appendix~\ref{appendix:proof of power saving for DLT NOMA over OMA-II}
\end{IEEEproof}

Then, assuming \(\{p_1^\ast(\nu),p_2^\ast(\nu),\alpha_1^\ast(\nu)\}\) as the optimal solution to \(\mathrm{(P2\text{-}OMA\text{-}II)}\), we construct a solution to \(\mathrm{(P2\text{-}NOMA)}\) based on \(\{p_1^\ast(\nu),p_2^\ast(\nu)\}\) as follows. We need to modify solution (c.f.~\eqref{eq:case-specified optimal power allocation for DLT OMA-II under full CSIT}) to \(\mathrm{(P2\text{-}OMA\text{-}II\text{-}sub)}\) corresponding to each of the four cases. It is straightforward to check that for the first three cases in~\eqref{eq:case-specified optimal power allocation for DLT OMA-II under full CSIT}, setting \(p_1=p_1^\ast\) and \(p_2=p_2^\ast\) corresponds to the first three cases in~\eqref{eq:case-specified optimal power allocation for DLT NOMA under full CSIT}, and thus \(X_k^{\rm NOMA}=X_k^{\ast\rm OMA\text{-}II}\), \(\forall k\). If \((p_1^\ast,p_2^\ast)\) falls in the last case of~\eqref{eq:case-specified optimal power allocation for DLT OMA-II under full CSIT}, we reallocate them among the two users such as \(p_1=\frac{2^{\bar R_1}-1}{g_1}\) and \(p_2=(2^{\bar R_{2}}-1)(\frac{2^{\bar R_1}-1}{g_1}+\frac{1}{g_{2}})\) assuming \(g_1>g_2\) w.l.o.g., and therefore it follows that \(X_k^{\rm NOMA}=X_k^{\ast\rm OMA\text{-}II}=1\), \(\forall k\). The modification is feasible, since \(p_1^\ast+p_2^\ast\ge p_1+p_2\). This is because when $K=2$, it follows from Lemma~\ref{lemma:power saving for DLT NOMA over OMA-II} that
\begin{align*}
\min\limits_{\sum_{i=1}^2\alpha_k=1}\sum_{i=1}^2\frac{(2^{\frac{\bar R_i}{\alpha_i}}-1)\alpha_i}{g_i}\ge \frac{2^{\bar R_2}-1}{g_2}+\frac{2^{\bar R_2}(2^{\bar R_1}-1)}{g_1},
\end{align*} 
in which the left-hand side (LHS) and right right-hand side (RHS) corresponds to \(p_1^\ast+p_2^\ast\) and \(p_1+p_2\), respectively.

To sum up, with the constructed solution \((p_1,p_2)\) in each fading state, it follows that \(\mb{E}_\nu[X_k^{\rm NOMA}(\nu)]=\mb{E}_\nu[X_k^{\ast\rm OMA\text{-}II}(\nu)]\le\bar\zeta\), \(\forall k\). Moreover, benefiting from the saved power by NOMA, \eqref{eq:APC under full CSIT} now becomes inactive, which suggests that \(\mb{E}_\nu[X_k^{\rm NOMA}(\nu)]\)'s are potentially to be further reduced should the power be fully allocated. Hence, \(\mathrm{(P2\text{-}NOMA)}\) is shown to be able to achieve larger sum of DLT than \(\mathrm{(P2\text{-}OMA\text{-}II)}\).

\subsection{Partial CSIT}
Problem \(\mathrm{(P2\text{-}XX)}\) under partial CSIT is recast as follows:
\begin{subequations}
\begin{align}
\mathrm{(P2^\prime\text{-}XX)}:~\mathop{\mathtt{Maximize}}_{p_s,p_w,\alpha_k}
& ~~~ \bar R_k(1-\zeta_k^{\prime\rm XX})+\bar R_{\bar k}(1-\zeta_{\bar k}^{\prime\rm XX})\notag\\
\mathtt {Subject \ to}& ~~~p_s+p_w\leq \bar P,\\
 & ~~~p_s\ge 0, \ p_w\ge 0,\\
 & ~~~\mb{E}_\nu[X_k^{\prime\rm XX}(\nu)]\le\bar\zeta^\prime, \; \forall k,\label{eq:max outage constraint under partial CSIT} 
\end{align}
\end{subequations} 
where \(\alpha_k\)'s is only valid when ``XX'' is replaced by OMA-Type-II, and \eqref{eq:max outage constraint under partial CSIT} constrain the maximum user outage probability of the two below \(\bar\zeta^\prime\). We present in the sequel how to optimally solve \(\mathrm{(P2^\prime\text{-}NOMA)}\) and \(\mathrm{(P2^\prime\text{-}OMA\text{-}II)}\), respectively. 

\subsubsection{Optimal Solution to \(\mathrm{(P2^\prime\text{-}NOMA)}\)}\label{subsubsec:Optimal Solution to (P2'-NOMA) under Partial CSIT}
In line with the same notation for channel gains as defined in \eqref{eq:conditional CDF of Gamma_k in NOMA} and \eqref{eq:conditional CDF of tilde Gamma_k in NOMA}, replace \(p_k(\nu)\) with \(p_s\), \(p_{\bar k}(\nu)\) with \(p_w\), \(\forall\nu\), when \(X>Y\), and otherwise do this reversely. As a result, \(\mb{E}_\nu[X_k^{\prime\rm NOMA}(\nu)]=\zeta_k^{\prime\rm NOMA}\) can be recast as follows:
\begin{multline}
\mb{E}_\nu[X_k^{\prime\rm NOMA}(\nu)]=\Pr\left\{\log_2\left(1+\frac{p_sX}{p_wX+\sigma_k^2} \right )<\bar R_k,\right.\\
\left.\log_2\left(1+\frac{p_wX}{p_sX+\sigma_k^2} \right )<\bar R_{\bar k}, X>Y\right\}
+\Pr\bigg\{\\
\log_2\left(1+\frac{p_wX}{p_sX+\sigma_k^2} \right )\ge\bar R_{\bar k}, X>Y \bigg\}+\\
\Pr\left\{\log_2\left(1+\frac{p_wX}{p_sX+\sigma_k^2}\right)<\bar R_k, X\le Y\right\}. \label{eq:sigma_k prime NOMA under partial CSIT}
\end{multline}
With CDI regarding \(X\) and \(Y\) given in Section~\ref{sec:System Model}, \(\mb{E}_\nu[X_k^{\prime\rm NOMA}(\nu)]\)'s can be derived based upon \eqref{eq:sigma_k prime NOMA under partial CSIT} shown in the following proposition.
\begin{proposition}   
The outage probability for NOMA user \(\mathcal{U}_k\) given the prescribed transmit rate \(\bar R_k\), \(k\in\{1,2\}\), under partial CSIT is given by
\begin{align}
&\mb{E}_\nu[X_k^{\prime\rm NOMA}(\nu)]=\notag\\
&\kern-8pt\left\{\begin{array}{l} \kern-4pt 1-e^{-\lambda_k\varepsilon_{k,j}}+\frac{\lambda_k}{\lambda_k+\lambda_{\bar k}}e^{-(\lambda_k+\lambda_{\bar k})\varepsilon_{k,j}}-\\
\frac{\lambda_k}{\lambda_k+\lambda_{\bar k}}e^{-(\lambda_k+\lambda_{\bar k})\varepsilon_{k,3}},\ {\bf(a)} \\
\kern-4pt 1-e^{-\lambda_k\varepsilon_{k,j}}+\frac{\lambda_k}{\lambda_k+\lambda_{\bar k}}e^{-(\lambda_k+\lambda_{\bar k})\varepsilon_{k,j}}, \ {\bf(b)}\\
\kern-4pt 1-\frac{\lambda_k}{\lambda_k+\lambda_{\bar k}}e^{-(\lambda_k+\lambda_{\bar k})\varepsilon_{k,3}}, \ {\bf(c)}\\
\kern-4pt \frac{\lambda_{\bar k}}{\lambda_k+\lambda_{\bar k}}-e^{-\lambda_k\varepsilon_{k,1}}+\frac{\lambda_k}{\lambda_k+\lambda_{\bar k}}e^{-(\lambda_k+\lambda_{\bar k})\varepsilon_{k,1}}, \ {\bf(d)} \\
\kern-4pt 1, \ {\bf(e)}
\end{array}\right. \label{eq:case-specific zeta_k prime NOMA under partial CSIT}
\end{align}
where \(\varepsilon_{k,1}\triangleq\frac{\sigma_k^2\tau_k}{p_s}\), \(\varepsilon_{k,2}\triangleq\frac{\sigma_k^2\tau_k}{p_s-p_w\tau_k}\), \(\varepsilon_{k,3}\triangleq\frac{\sigma_k^2\tau_k}{p_w-p_s\tau_k}\), and \(\varepsilon_{k,4}\triangleq\frac{\sigma_k^2\tau_{\bar k}}{p_w-p_s\tau_{\bar k}}\), with \(\tau_k\triangleq 2^{\bar R_k}-1\). The conditions in (a)-(e) of \eqref{eq:case-specific zeta_k prime NOMA under partial CSIT} corresponds to
\begin{align}
\left\{\begin{array}{l}
\alpha_{k,c_j^1}\le\frac{p_s}{p_w}\le\beta_{k,c_j^1},\; j\in\{1,2,3\},\\
\alpha_{k,c_j^2}\le\frac{p_s}{p_w}\le\beta_{k,c_j^2},\; j\in\{1,2,3\},\\
\alpha_{k,c^3}\le\frac{p_s}{p_w}\le\beta_{k,c^3},\\
\alpha_{k,c^4}\le\frac{p_s}{p_w}\le\beta_{k,c^4},\\
\alpha_{k,c^5}\le\frac{p_s}{p_w}\le\beta_{k,c^5},
\end{array}\right. \label{eq:case-specific alpha and beta under partial CSIT}
\end{align} respectively.
In \eqref{eq:case-specific alpha and beta under partial CSIT}, \(\alpha_{k,c_j^1}\) and \(\beta_{k,c_j^1}\), \(j\in\{1,2,3\}\), are given by\footnote{The parameter values preceding and coming after ``or'' form a pair, respectively.}
{\begin{align*}
\left\{\begin{array}{l} 
\alpha_{k,c_1^1}=0\\ 
\alpha_{k,c_2^1}=\max\left\{\frac{\tau_k(\tau_{\bar k}+1)}{\tau_{\bar k}(\tau_k+1)},\tau_k\right\}\, \mbox{or}\, \max\left\{\tau_k,\frac{1}{\tau_{\bar k}}\right\}\\
\alpha_{k,c_3^1}=\max\left\{\frac{\tau_k}{\tau_{\bar k}(\tau_k+1)},\tau_k\right\}\, \mbox{or}\, \frac{\tau_k}{\tau_{\bar k}(\tau_k+1)}
\end{array}\right.
%\label{eq:alpha_k and beta_k for the first case}
\end{align*}} and
\begin{align*}
\left\{\begin{array}{l}
\beta_{k,c_1^1}=\min\left\{\frac{\tau_k}{\tau_{\bar k}(\tau_k+1)},\frac{1}{\tau_k}\right\}\\
\beta_{k,c_2^1}=\min\left\{\frac{1}{\tau_k},\frac{1}{\tau_{\bar k}}\right\}\, \mbox{or}\, \frac{1}{\tau_k}\\
\beta_{k,c_3^1}=\min\left\{\frac{\tau_k(\tau_{\bar k}+1)}{\tau_{\bar k}(\tau_k+1)},\frac{1}{\tau_k},\frac{1}{\tau_{\bar k}}\right\}\, \mbox{or}\, \min\left\{\tau_k,\frac{1}{\tau_k},\frac{1}{\tau_{\bar k}}\right\}
\end{array}.\right.
\end{align*} 
Moreover, \(\alpha_{k,c_j^2}\) and \(\beta_{k,c_j^2}\), \(j\in\{1,2,3\}\), are given by
\begin{align*}
\kern-4pt\left\{\begin{array}{l} 
\alpha_{k,c_1^2}=\max\left\{\tau_k,\frac{1}{\tau_k}\right\}\\ 
\alpha_{k,c_2^2}=\max\left\{\frac{\tau_k(\tau_{\bar k}+1)}{\tau_{\bar k}(\tau_k+1)},\tau_k,\frac{1}{\tau_k}\right\}\, \mbox{or}\, \max\left\{\tau_k,\frac{1}{\tau_k},\frac{1}{\tau_{\bar k}}\right\}\\
\alpha_{k,c_3^2}=\max\left\{\frac{\tau_k}{\tau_{\bar k}(\tau_k+1)},\tau_k,\frac{1}{\tau_k}\right\}\, \mbox{or}\, \max\left\{\frac{\tau_k}{\tau_{\bar k}(\tau_k+1)},\frac{1}{\tau_k}\right\}    
\end{array}\right.
%\label{eq:alpha_k and beta_k for the second case}
\end{align*} and
\begin{align*}
\left\{\begin{array}{l}
\beta_{k,c_1^2}=\frac{\tau_k}{\tau_{\bar k}(\tau_k+1)}\\
\beta_{k,c_2^2}=\frac{1}{\tau_{\bar k}}\, \mbox{or}\, +\infty\\
\beta_{k,c_3^2}=\min\left\{\frac{\tau_k(\tau_{\bar k}+1)}{\tau_{\bar k}(\tau_k+1)},\frac{1}{\tau_{\bar k}}\right\}\, \mbox{or}\, \min\left\{\tau_k,\frac{1}{\tau_{\bar k}}\right\}
\end{array}. \right.
\end{align*} 
Finally, \(\alpha_{k,c^l}\) and \(\beta_{k,c^l}\), \(l\in\{3,4,5\}\), are given by
\begin{align*}
\kern-4pt\left\{\begin{array}{l} 
\alpha_{k,c^3}=\frac{1}{\tau_{\bar k}}\\  
\alpha_{k,c^4}=\frac{1}{\tau_k}\\  
\alpha_{k,c^5}=\max\left\{\frac{1}{\tau_k},\frac{1}{\tau_{\bar k}}\right\}  
\end{array},\right.\kern-1pt 
\left\{\begin{array}{l}
\beta_{k,c^3}=\min\left\{\tau_k,\frac{1}{\tau_k}\right\}\\ 
\beta_{k,c^4}=\min\left\{\frac{\tau_k}{\tau_{\bar k}(\tau_k+1)},\tau_k\right\}\\
\beta_{k,c^5}=\tau_k
\end{array}.\right.
%\label{eq:alpha_k and beta_k for the third to the fifth case}
\end{align*} \label{prop:case-specific sigma_k prime NOMA under partial CSIT}
\end{proposition} 
\begin{IEEEproof}
Please refer to Appendix~\ref{appendix:proof of case-specific sigma_k prime NOMA under partial CSIT}.
\end{IEEEproof}

In accordance with Proposition~\ref{prop:case-specific sigma_k prime NOMA under partial CSIT}, we are able to derive the sum of the DLT \(\bar R_k(1-\zeta_k^{\prime\rm NOMA})+\bar R_{\bar k}(1-\zeta_{\bar k}^{\prime\rm NOMA})\). Problem \(\mathrm{(P2^\prime\text{-}NOMA)}\) can also be characterized by only one optimization variable \(p_s\) by replacing \(p_w\) with \(\bar P-p_s\), and then optimally solved (up to numerical accuracy) via one-dimension search over \(p_s\in[0,\bar P]\).

\subsubsection{Optimal Solution to \(\mathrm{(P2^\prime\text{-}OMA\text{-}II)}\)}\label{subsubsec:Optimal Solution to (P2'-OMA-II) under Partial CSIT}
In line with the principle of power and time/frequency allocations for OMA-Type-II transmission described above ~\eqref{eq:R_k prime OMA-II under partial CSIT}, replace \(p_k(\nu)\) with \(p_s\), \(p_{\bar k}(\nu)\) with \(p_w\), \(\forall\nu\), when \(X>Y\), and the reverse when \(X\le Y\) in \eqref{eq:sigma_k OMA-II under full CSIT}. \(\zeta_k^{\prime\rm OMA\text{-}II}=\mb{E}_\nu[X_k^{\prime\rm OMA\text{-}II}(\nu)]\), is derived as follows:
{\begin{multline}
\mb{E}_\nu[X_k^{\prime\rm OMA\text{-}II}(\nu)]=\\
\Pr\left\{\alpha_k\log_2\left(1+\frac{p_sX}{\alpha_k\sigma_k^2} \right )<\bar R_k, X>Y\right\}+\\
\Pr\left\{\alpha_k\log_2\left(1+\frac{p_wX}{\alpha_k\sigma_k^2} \right )<\bar R_{\bar k},X\le Y\right\}. \label{eq:sigma_k prime OMA-II under partial CSIT}
\end{multline}} We are thus able to derive \(\mb{E}_\nu[X_k^{\prime\rm OMA\text{-}II}(\nu)]\) in the following proposition.
\begin{proposition}  
The outage probability for OMA-Type-II user \(\mathcal{U}_k\) given the prescribed transmit rate \(\bar R_k\), \(k\in\{1,2\}\), under partial CSIT is given by
\begin{multline}
\mb{E}_\nu[X_k^{\prime\rm OMA\text{-}II}(\nu)]=1+\frac{\lambda_k}{\lambda_k+\lambda_{\bar k}}\big(e^{(\lambda_k+\lambda_{\bar k})\varphi_{k,1}}-\\
e^{(\lambda_k+\lambda_{\bar k})\varphi_{k,2}}\big)-
e^{-\lambda_k\varphi_{k,1}}, \label{eq:analytical sigma_k prime OMA-II under partial CSIT}
\end{multline} where \(\varphi_{k,1}\triangleq\frac{\alpha_k\sigma_k^2\xi_k}{p_s}\) and \(\varphi_{k,2}\triangleq\frac{\alpha_k\sigma_k^2\xi_k}{p_w}\), with \(\xi_k\triangleq 2^{\frac{\bar R_k}{\alpha_k}}-1\).
\label{prop:analytical sigma_k prime OMA-II under partial CSIT}
\end{proposition} 
\begin{IEEEproof}
With CDI of \(X\) and \(Y\) known, the derivation of \(\mb{E}_\nu[X_k^{\prime\rm OMA\text{-}II}(\nu)]\) from \eqref{eq:sigma_k prime OMA-II under partial CSIT} is straightforward and thus omitted here for brevity.  
\end{IEEEproof}

Based on Proposition~\ref{prop:analytical sigma_k prime OMA-II under partial CSIT}, \(\mathrm{(P2^\prime\text{-}OMA\text{-}II)}\) can be solved similarly as \(\mathrm{(P2^\prime\text{-}NOMA)}\), the detail of which is omitted herein for brevity.

\section{Numerical Results}\label{sec:Numerical Results}
In this section, we verify the theoretical analysis for the considered two-user downlink NOMA system via numerical results. \textcolor{black}{As a performance bench mark, we also provide one classical type of OMA transmission scheme, referred as {\em OMA-Type-I}, which assigns equal amount of time (in TDMA) or frequency (in FDMA) resources among users over all fading states, i.e.,  \(\alpha_k(\nu)=\frac{1}{2}\), \(\forall k\), \(\forall\nu\) in \eqref{eq:R_k OMA-II under full CSIT}, and \(\alpha_k=\frac{1}{2}\), \(\forall k\), in \eqref{eq:R_k prime OMA-II under partial CSIT}.} The corresponding optimal power policies to OMA-Type-I are easily seen to be special cases of  OMA-Type-II, which has already been solved. 
\(\mathcal{U}_1\) and \(\mathcal{U}_2\) are assumed to be located with a distance of \(d_1\) and \(d_2\) away from the BS, respectively. The large-scale path loss model of the channel is given by \(128.1+37.6\log10(D)\) in dB, where \(D\) in kilometer (km) denotes the distance from the BS to the user. The small-scale fading is assumed to be  independent and identically distributed ($i.i.d.$) Rayleigh fading. The AWGNs at the users' Rxs are both assumed to be \(-169\)dBm/Hz over \(10\)MHz bandwidth. The infinite number of fading states is approximated by \(10^7\). Other simulation parameters are set as follows: \(d_1=0.1\)km, \(d_2=0.5\)km, \(\hat P=5\)Watt and \(\bar P=1\)Watt unless otherwise specified. 
\subsection{Delay-Tolerant Transmission}
\begin{figure}[htp]
\centering
\includegraphics[width=3.0in]{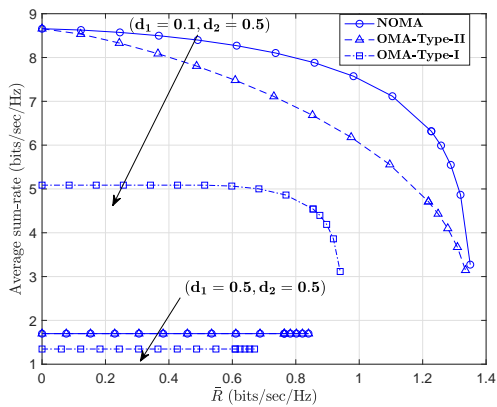}
\caption{The average sum-rate of the system versus the minimum rate constraints under full CSIT.}\label{fig:ESR under full CSIT}
\end{figure}

Fig.~\ref{fig:ESR under full CSIT} depicts the optimal trade-offs between the average sum-rate of the system and the minimum rate constraints under full CSIT, i.e., \(\bar R\), achieved by NOMA and the OMA schemes with different distance settings. 
It is seen that with the near-far distance setting, NOMA outperforms OMA-Type-II transmission in most cases, while the gap shrinks when \(\bar R\) is very little and/or approaches \(\bar R_{\max}\), respectively. Moreover, both NOMA and OMA-Type-II achieve substantially larger optimal trade-off than OMA-Type-I, although OMA-Type-I is seen more robust against increase in \(\bar R\). This is because OMA-Type-I is intrinsically of fairness in view of equal time/frequency assigned to each user irrespective of their CSI.
It is also worth noting that when there is no difference between the two users in terms of large-scale fading, the average sum-rate versus min-rate trade-offs almost vanish, since the average sum-rate w/o the minimum rate constraint has already achieved certain fairness, i.e., \(\mb E_\nu[R_k^{\rm NOMA}(\nu)]\approx\mb E_\nu[R_{\bar k}^{\rm NOMA}(\nu)]\) (\(\mb{E}_\nu[R_k^{\rm OMA\text{-}II}(\nu)]\approx\mb{E}_\nu[R_{\bar k}^{\rm OMA\text{-}II}(\nu)]\)) due to their statistically similar channel distribution.
%Furthermore, the average sum-rate achieved by NOMA is seen to be achieved by OMA-Type-II with negligible gap, since in this case, the achievable average sum-rate for both schemes remains approximately the same as that w/o minimum average constraint, the latter of which achieved by OMA-Type-II is in theory the same as that achieved by NOMA. 

\begin{figure}[htp]
\centering
\includegraphics[width=3.0in]{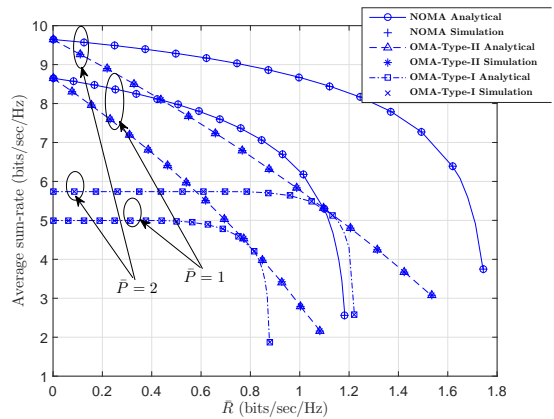}
\caption{The average sum-rate of the system versus the minimum rate constraints under partial CSIT.}\label{fig:ESR under partial CSIT}
\end{figure}

Fig.~\ref{fig:ESR under partial CSIT} shows the optimal trade-offs between the average sum-rate of the system versus the minimum rate constraints under partial CSIT, i.e., \(\bar R^\prime\), achieved by various multiple access schemes with different APC. The optimal trade-off regions between the average-sum rate of the system and the fairness are expectedly seen to enlarge with increasing limit on the transmit power \(\bar P\). While the superiority of the proposed power allocation policies for NOMA against OMA-Type-II is obviously seen, the contrast is more sharply observed for NOMA against OMA-Type-I in Fig. ~\ref{fig:ESR under full CSIT}.

\begin{figure}[htp]
\centering
\subfigure[Full CSIT. \label{subfig:ESR user fairness under full CSIT}]{\includegraphics[width=3.0in]{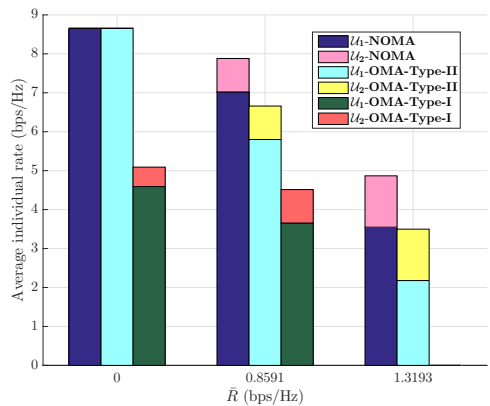}}
\subfigure[Partial CSIT.\label{subfig:ESR user fairness under partial CSIT}]{\includegraphics[width=3.0in]{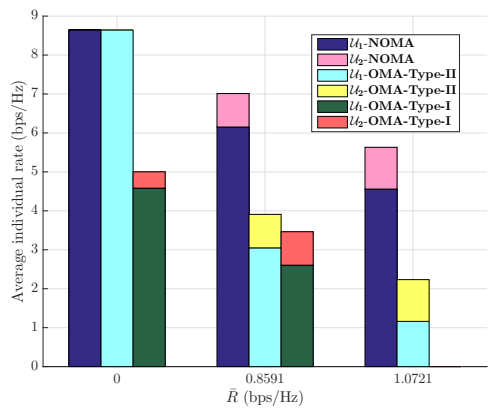}}
\caption{The average achievable rate allocation between the two users with different minimum rate requirements, under full and partial CSIT, respectively.}\label{fig:ESR user fairness}
\end{figure}

The comparison between the individual ergodic rate subject to varied minimum average rate constraints is demonstrated in Fig. \ref{subfig:ESR user fairness under full CSIT} (Fig. \ref{subfig:ESR user fairness under partial CSIT}) for NOMA, OMA-Type-I, and OMA-Type-II, respectively, under full (partial) CSIT. First, we see that OMA-Type-II achieves almost the same ergodic rate for \(\mathcal{U}_1\) as NOMA with \(\mathcal{U}_2\)'s ergodic rate both as little as zero, when there is no minimum rate requirement. This can be intuitively explained as follows. Since \(\mathcal{U}_1\) is the near user who enjoys better CSI in most of the fading states, the optimal power policy that maximizes the average sum-rate for both NOMA and OMA-Type-II is to allocate power only to \(\mathcal{U}_1\) in such states. Moreover, the advantage of NOMA begins promising when the system requires a larger \(\bar R\) (\(\bar  R^\prime\)), in that NOMA guarantees the minimum average rate achieved by \(\mathcal{U}_2\) while keeping \(\mathcal{U}_1\)'s average rate the maximum. 
%Furthermore, owing to complete knowledge of CSIT, the maximum permissive min-rate in the full CSIT case, \(\bar R_{\max}\approx1.3193\)b/Hz, is larger than \(\bar R_{\max}\approx1.0721\) in the partial CSIT case.

\subsection{Delay-Limited Transmission}
\begin{figure}[htp]
\centering
\subfigure[\label{subfig:DLT near far under full CSIT}]{\includegraphics[width=3.0in]{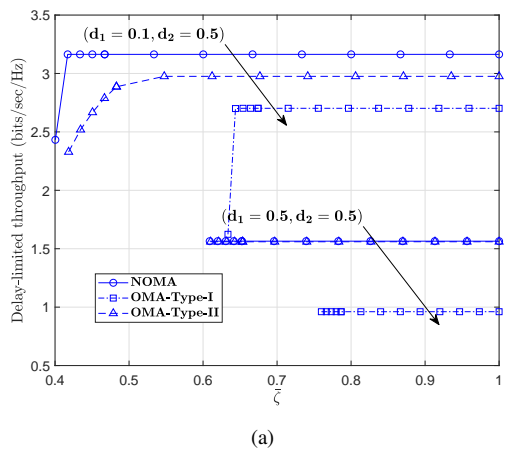}}
\subfigure[\label{subfig:DLT varied bar_R under full CSIT}]{\includegraphics[width=3.0in]{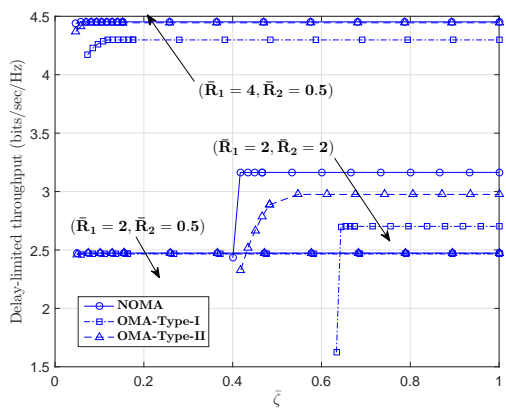}}
\caption{The DLT of the system versus the maximum permissive outage probability under full CSIT.}\label{fig:DLT under full CSIT}
\end{figure}

Fig. \ref{subfig:DLT near far under full CSIT} shows the optimal trade-offs between the sum of DLT and the maximum permissive outage probability, i.e., \(\bar\zeta\), under full CSIT given the same prescribed rate \(\bar R_k=2\)bits/sec/Hz for each user. It is seen that when the two users suffer from near-far unfairness, the optimum sum-DLT versus max-outage trade-off achieved by NOMA outperforms that achieved by OMA-Type-II and OMA-Type-I. However, this superiority almost disappears when \(\mathcal{U}_1\) and \(\mathcal{U}_2\) are both \(0.5\)km away from the BS.  This is because in this case \(g_1(\nu)\approx g_2(\nu)\) in most fading states, thanks to which the total amount of transmit power saved by NOMA tends to be less. Furthermore, no much trade-off is seen for the sum of DLT versus user fairness, as the two users hold similar chances to be the stronger user, and therefore when \(\zeta_k^{\rm XX}\) is minimized, \(\zeta_{\bar k}^{\rm XX}\) is nearly minimized as well, where \((\cdot)^{\rm XX}\) stands for \({\rm NOMA}\) or \({\rm OMA\text{-}Type\text{-}II}\). 

On the other hand, in near-far channel conditions, the impact of different \(\bar R_k\)s on the optimum sum-DLT versus max-outage trade-off is demonstrated in Fig. \ref{subfig:DLT varied bar_R under full CSIT}. With the same intended rate \(\bar R_1=\bar R_2=2\)bits/sec/Hz, the optimum trade-off achieved by NOMA outperforms that achieved by the OMA schemes. By contrast, when \(\bar R_2\) reduces to \(0.5\)bits/sec/Hz, the trade-off becomes trivial, since in this case the stronger user's advantage in saving power is compromised by its higher target rate. 
%When \(\bar R_1\) continues increasing to \(4\)bits/sec/Hz, the trade-off is seen come back with superior sum of DLT over that in other settings of \(\bar R_1\) and \(\bar R_2\).

\begin{figure}[htp]
\centering
%\subfigure[\label{subfig:DLT varied bar_P under partial CSIT}]{\includegraphics[width=2.8in]{eps/DLT_vs_max_outage_varied_bar_P_imperfect.eps}}
\includegraphics[width=3.0in]{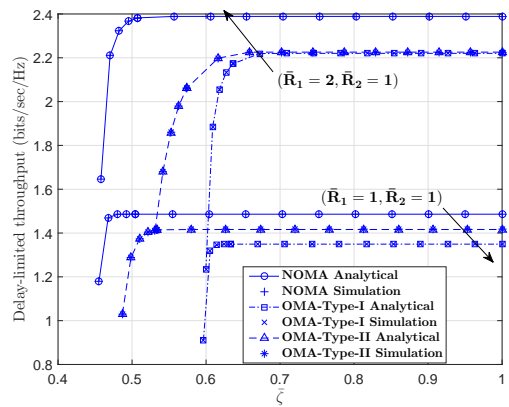}
\caption{The DLT of the system versus the maximum permissive user outage probability under partial CSIT.}\label{fig:DLT varied bar_R under partial CSIT}
\end{figure}
Fig. \ref{fig:DLT varied bar_R under partial CSIT} shows the optimum sum-DLT versus max-outage trade-offs achieved by various schemes with different settings of \(\bar R_1\) and \(\bar R_2\). Unlike in Fig. \ref{subfig:DLT varied bar_R under full CSIT}, the superiority of NOMA over the other OMA schemes is significantly seen in Fig. \ref{fig:DLT varied bar_R under partial CSIT}. Moreover, the minimum max-outage achieved by NOMA is significantly lower than that attained by other schemes. For example, with \(\bar R_1=\bar R_2=1\)bist/sec/Hz, \(\min\limits_{p_s,p_w}\{\max\limits_{k}\zeta_k^{\prime\rm NOMA}\}\) falls below \(0.47\) while that achieved by OMA-Type-I and OMA-Type-II is as large as about \(0.49\) and \(0.60\), respectively.

\begin{figure}[htp]
\centering
\subfigure[Full CSIT. \label{subfig:DLT user fairness under full CSIT}]{\includegraphics[width=3.0in]{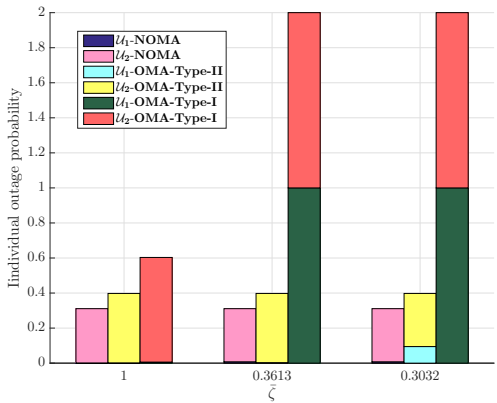}}
\subfigure[Partial CSIT.\label{subfig:DLT user fairness under partial CSIT}]{\includegraphics[width=3.0in]{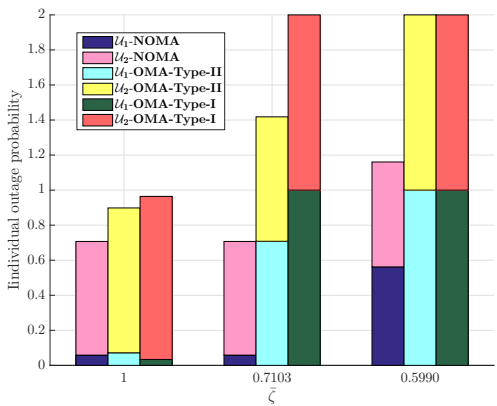}}
\caption{The DLT allocation between the two users with different maximum outage requirements, under full and partial CSIT.}\label{fig:DLT user fairness}
\end{figure}
%When \(\bar R_k\)s are set equal to each other, the individually achievable DLT under full (partial) CSIT, namely, \(\bar R_k(1-\zeta_k^{\rm XX})\) (\(\bar R_k(1-\zeta_k^{\prime\rm XX})\)), can be equivalently characterized by their individual outage probability \(\zeta_k^{\rm XX}\) (\(\zeta_k^{\prime\rm XX}\)). 
The DLT allocation between the two users subject to different maximum permissive outage is reflected by their outage probability allocation in Fig. \ref{fig:DLT user fairness} under full and partial CSIT, respectively. Under full CSIT, it is seen from Fig.~\ref{subfig:DLT user fairness under full CSIT} that with \(\bar R_1=\bar R_2=2\)bits/sec/Hz and \(\bar P=2\)Watt, \(\mathcal{U}_1\) achieves almost negligible outage while \(\mathcal{U}_2\) can achieve an outage probability as low as \(0.3032\) by NOMA. OMA-Type-II follows the same trend unless \(\mathcal{U}_1\) has to claim more outage states to reserve power for \(\mathcal{U}_2\)'s transmission. \(\mathcal{U}_1\)'s outage probability compromised by satisfying a lower maximum outage constraint is larger in the case of partial CSIT than in the case of full CSIT.

\section{Conclusion}\label{sec:Conclusion}
In this paper, we have investigated the average sum-rate and/or the sum of DLT maximization for a two-user downlink NOMA over fading channels imposing QoS constraints on the worst user performance. Under full CSIT, the non-convex resource allocation problems have been solved using the technique of dual decomposition leveraging ``time-sharing'' conditions. Under partial CSIT, the individual ergodic rate and/or outage probability have been characterized in closed-form, based on which the optimal power policies have been numerically obtained. Simulation results have unveiled that the optimal NOMA-based power allocation schemes in general outperform the optimal OMA-based ones in terms of various throughput versus fairness trade-offs, especially when the two users' channels experience contrasting fading gains. 

\begin{appendices}
\section{}\label{appendix:proof of p1+p2=hat P in OMA-II}
As \(p_1+p_2=\hat P\) is fixed,  \(\mathrm{(P1\text{-}OMA\text{-}II\text{-}sub)}\) reduces  to the following problem irrespective of \(\lambda\):
\begin{subequations}
\begin{align}
\mathop{\mathtt{Maximize}}_{p_1\ge 0,p_2\ge 0,\alpha_1}
& ~~(1+\delta)\alpha_1\log_2\left(1+\frac{p_1g_1}{\alpha_1}\right)+\notag\\
&~~(1+\mu)\alpha_2\log_2\left(1+\frac{p_2g_2}{\alpha_2}\right)\notag	\\
\mathtt {Subject \ to}& ~~p_1+p_2\le\hat P, \label{eq:modified active PPC}\\
&~~0\le\alpha_1\le 1,
\end{align} \label{eq:reduced problem of (P1-OMA-II-sub)}
\end{subequations} where \(\alpha_2=1-\alpha_1\). Note that \(=\) has been relaxed into \(\le\) in~\eqref{eq:modified active PPC}, since it is easy to check that the above problem obtains its optimum value when \eqref{eq:modified active PPC} is active. 
	
To facilitate solving \eqref{eq:reduced problem of (P1-OMA-II-sub)}, we introduce a new Lagrangian multiplier \(\lambda^\prime\) associated with the constraint \eqref{eq:modified active PPC}. Then given \(\lambda^\prime\), we aim for maximizing the Lagrangian regarding Problem \eqref{eq:reduced problem of (P1-OMA-II-sub)} as follows: 
\begin{subequations} 
\begin{align}
\mathop{\mathtt{Maximize}}_{p_1\ge 0,p_2\ge 0,\alpha_1}
& ~~~\bar{\mathcal{L}}_1^{\prime\rm OMA\text{-}II}(p_1,p_2,\alpha_1,\lambda^\prime)\notag\\
\mathtt {Subject \ to}& ~~~0\le\alpha_1\le 1,
\end{align}\label{eq:maximizing Lagrangian of (P1-OMA-II-sub)}
\end{subequations}
in which \(\bar{\mathcal{L}}_1^{\prime\rm OMA\text{-}II}(p_1,p_2,\alpha_1,\lambda^\prime)=(1+\delta)\alpha_1\log_2(1+\frac{p_1g_1}{\alpha_1})
+(1+\mu)(1-\alpha_1)\log_2(1+\frac{p_2g_2}{1-\alpha_1})-\lambda^\prime(p_1+p_2)\). It is worthy of noting that Problem \eqref{eq:maximizing Lagrangian of (P1-OMA-II-sub)} shares the same structure as \(\mathrm{(P1\text{-}OMA\text{-}II\text{-}sub)}\) except that the constraint \(p_1+p_2\le\hat P\) is now removed.  Therefore, the rationale behind solving it has been similarly given in the proof for Proposition~\ref{prop:optimal power allocation for ESR OMA-II}. As a result, given \(\lambda^\prime\), the maximizer of \(\bar{\mathcal{L}}_1^{\prime\rm OMA\text{-}II}(p_1,p_2,\alpha_1,\lambda^\prime)\) proves to be either \((0,(\frac{1+\mu}{\lambda^\prime\ln 2}-\frac{1}{g_2})^+,0)\) or \(((\frac{1+\delta}{\lambda^\prime\ln 2}-\frac{1}{g_1})^+,0,1)\), i.e., alternative transmission. Therefore, by updating \(\lambda^\prime\) via bi-section until \eqref{eq:modified active PPC} is active, the optimal solution to Problem \eqref{eq:reduced problem of (P1-OMA-II-sub)} ends up with \((0,\hat P,0)\) or \((\hat P,0,1)\) depending on which leads to a larger objective value. Hence, we complete the proof for Lemma~\ref{lemma:p1+p2=hat P in OMA-II}.

\section{}\label{appendix:proof of optimal power allocation for ESR OMA-II}
As the maximum of \(\bar{\mathcal{L}}_1^{\rm OMA\text{-}II}(p_1,p_2,\alpha_1)\) must be either at the stationary point or on the boundary of \(\Psi\) as defined in Proposition~\ref{prop:optimal power allocation for ESR NOMA}, the vertexes of \((0,0,0)\), \((0,\hat P,0)\), and \((\hat P,0,1)\) are included in \eqref{eq:optimal power allocation for ESR OMA-II} for sure. 

As for another case that the maximum of \(\bar{\mathcal{L}}_1^{\rm OMA\text{-}II}(p_1,p_2,\alpha_1)\) is achieved on \(p_1=0\), the corresponding optimum value of \(\bar{\mathcal{L}}_1^{\rm OMA\text{-}II}(p_1,p_2,\alpha_1)\) takes on \(\bar{\mathcal{L}}_1^{\rm OMA\text{-}II}(0,c_2,0)\mathbbm{1}_{c_2}\) because of the following reasons. Plugging \(p_1=0\) into \(\bar{\mathcal{L}}_1^{\rm OMA\text{-}II}(p_1,p_2,\alpha_1)\), its derivative w.r.t \(p_2\) and \(\alpha_1\) are, respectively, expressed as:
\begin{align}
\frac{\partial\bar{\mathcal{L}}_1^{\rm OMA\text{-}II}(p_1,p_2,\alpha_1)}{p_2}  = & \frac{(1+\mu)\alpha_2g_2}{(\alpha_2+p_2g_2)\ln 2}-\lambda,
\label{eq:partial L1 partial p2 for OMA-II}\\
\frac{\partial\bar{\mathcal{L}}_1^{\rm OMA\text{-}II}(p_1,p_2,\alpha_1)}{\alpha_1} = & \frac{(1+\mu)}{\ln 2}\bigg[-\ln\left(1+\frac{p_2g_2}{1-\alpha_1} \right )+\notag\\
&1-\frac{1}{1+\Myfrac{p_2g_2}{(1-\alpha_1)}}\bigg]. \label{eq:partial L1 partial alpha1 for OMA-II}
\end{align} 
It is then easily seen that the optimal \(p_2\) admits the form of \(p_2^\ast=[c_2(1-\alpha_1)]_0^{\hat P}\). Further, \(\bar{\mathcal{L}}_1^{\rm OMA\text{-}II}(0,p_2,\alpha_1)\) turns out to monotonically decrease w.r.t \(\alpha_1\) by observing that the RHS of  \eqref{eq:partial L1 partial alpha1 for OMA-II} is always negative in view of the inequality \(1-\frac{1}{x}\le\ln x\). Hence, \(\alpha_1^\ast=0\), and \((0,c_2,0)\) is the optimum iff \(0\le c_2\le\hat P\), which leads to \(\bar{\mathcal{L}}_1^{\rm OMA\text{-}II}(0,c_2,0)\mathbbm{1}_{c_2}\). Similarly, \(\bar{\mathcal{L}}_1^{\rm OMA\text{-}II}(c_1,0,1)\mathbbm{1}_{c_1}\) can be justified considering another boundary of \(p_2=0\).

At last, the reasons why the jointly stationary point (c.f.~\eqref{eq:sp in OMA-II}) cannot be the optimal solution to \(\mathrm{(P1\text{-}OMA\text{-}II\text{-}sub)}\) has been explained in Remark~\ref{remark:at most one user transmit}.

\section{}\label{appendix:proof of ER for the kth user in NOMA}
\begin{align}
&\mb{E}_\nu[{R_k^{\prime\rm NOMA}(\nu)}]\notag\\ 
 &=  \Pr \{X \ge Y\}\mb{E}_\nu\left[ {\left. {{{\log }_2}\left( {1 + {\Gamma _{k}}} \right)} \right|X \ge Y} \right]+\notag\\
 &\ ~~  \Pr \{X < Y\}\mb{E}_\nu\left[ {\left. {{{\log }_2}\left( {1 + {\tilde\Gamma _{k}}} \right)} \right|X < Y} \right]\notag\\
 &= \frac{1}{{\ln 2}}\Pr \{X \ge Y\}\int_0^\infty  {\frac{{1 -{F_{{\Gamma _{k|X \ge Y}}}}\left( {{z}} \right)}}{{1 + z}}} \mathrm{d}z+\notag\\ 
 &\ ~~ \frac{1}{{\ln 2}}\Pr \{X < Y\}\int_0^\infty  {\frac{{1 - {F_{{\tilde\Gamma_k|X<Y }}}}\left( {{z}} \right) }{{1 + z}}} \mathrm{d}z\notag\\
&\stackrel{\rm (a)}{=}\frac{1}{{\ln 2}}\frac{{{\lambda _k}}}{{{\lambda _k} + {\lambda _{\bar k}}}}A_1 + \frac{1}{{\ln 2}}B_1 - \frac{1}{{\ln 2}}\frac{{{\lambda _k}}}{{{\lambda_k} + {\lambda _{\bar k}}}}C_1, \label{eq:calculate ER for the kth user in NOMA}
\end{align} where \(A_1\triangleq\int_0^{\frac{{{p_w}}}{{{p_s}}}} {\frac{{{e^{ - \left( {{\lambda _{k}} + {\lambda _{\bar k}}} \right)\frac{{\sigma _k^2z}}{{{p_w} - {p_s}z}}}}}}{{1 + z}}\mathrm{d}} z\), \(B_1\triangleq\int_0^\infty  {\frac{{{e^{ - \frac{{{\lambda _k}\sigma _k^2}}{{{p_s}}}z}}}}{{1 + z}}\mathrm{d}} z\), and \(C_1\triangleq\int_0^\infty  {\frac{{{e^{ - \frac{{\left( {{\lambda _k} + {\lambda _{\bar k}}} \right)\sigma _k^2}}{{{p_s}}}z}}}}{{1 + z}}\mathrm{d}} z\). Since \(\Pr\{X\ge Y\}=\frac{\lambda_{\bar k}}{\lambda_k+\lambda_{\bar k}}\), substituting \eqref{eq:conditional CDF of Gamma_k in NOMA} and \eqref{eq:conditional CDF of tilde Gamma_k in NOMA} for \(F_{\Gamma_k | X\ge Y}(z)\) and \(F_{\tilde\Gamma_k | X<Y}(z)\) in \eqref{eq:calculate ER for the kth user in NOMA}, respectively, \({\rm (a)}\) is derived. Then after some manipulations, by applying \cite[Eq.~(3.352.4)]{gradshteyn}, we have \(A_1= f(\frac{{( {{\lambda _k} + {\lambda _{\bar k}}} )\sigma _k^2}}{{{p_s}}})-f(\frac{{( {{\lambda _k} + {\lambda _{\bar k}}} )\sigma _k^2}}{{{p_s} + {p_w}}})\).
%\begin{align}
%A_1= f\left(\frac{{\left( {{\lambda _k} + {\lambda _{\bar k}}} \right)\sigma _k^2}}{{{p_s}}}\right)-f\left(\frac{{\left( {{\lambda _k} + {\lambda _{\bar k}}} \right)\sigma _k^2}}{{{p_s} + {p_w}}}\right). \label{eq:A1}
%\end{align} 
It also immediately follows that \(B_1=-f(\frac{{{\lambda _k}\sigma _k^2}}{{{p_s}}})\) and \(C_1=-f(\frac{{\left( {{\lambda _k} + {\lambda _{\bar k}}} \right)\sigma _k^2}}{{{p_s}}})\). Plugging \(A_1\), \(B_1\) and \(C_1\) into \eqref{eq:calculate ER for the kth user in NOMA}, Proposition~\ref{prop:ER for the kth user in NOMA} is proved.

\section{}\label{appendix:proof of ER for the kth user in OMA-II}
\begin{align}
&\mb{E}_\nu[R_k^{\prime\rm OMA\text{-}II}(\nu)]\notag \\
& =  \Pr \{X \ge Y\}\mb {E}_\nu\left[ {\left. \alpha_k{{{\log }_2}\left( {1 + {\Gamma _{k}}} \right)} \right|X \ge Y} \right]+\notag\\ 
&\ ~~\Pr \{X < Y\}\mb {E}_\nu\left[ {\left.\alpha_k {{{\log }_2}\left( {1 + {\tilde\Gamma _{k}}} \right)} \right|X < Y} \right]\notag\\
& =  \frac{\alpha_k}{{\ln 2}}\frac{\lambda_{\bar k}}{\lambda_k+\lambda_{\bar k}}\int_0^\infty  {\frac{{1 -{F_{{\Gamma _{k|X \ge Y}}}}\left( {{z}} \right)}}{{1 + z}}} \mathrm{d}z+\notag\\
&\ ~~\frac{\alpha_k}{{\ln 2}}\frac{\lambda_k}{\lambda_k+\lambda_{\bar k}}\int_0^\infty  {\frac{{1 - {F_{{\tilde\Gamma_k|X<Y }}}}\left( {{z}} \right) }{{1 + z}}} \mathrm{d}z\notag\\
& \stackrel{\rm (a)}{=} -\frac{\alpha_k}{{\ln 2}}\frac{{{\lambda _k}}}{{{\lambda _k} + {\lambda _{\bar k}}}}A_1^\prime + \frac{\alpha_k}{{\ln 2}}B_1^\prime +\frac{\alpha_k}{{\ln 2}}\frac{{{\lambda _k}}}{{{\lambda_k} + {\lambda _{\bar k}}}}C_1^\prime, \label{eq:calculate ER for the kth user in OMA-II}
\end{align} where \(A_1^\prime\triangleq\int_0^\infty  {\frac{{{e^{ - \frac{{\left( {{\lambda _k} + {\lambda _{\bar k}}} \right)\alpha_k\sigma _k^2}}{{{p_s}}}z}}}}{{1 + z}}\mathrm{d}} z\), \(B_1^\prime\triangleq\int_0^\infty  {\frac{{{e^{ - \frac{{{{\lambda _k} } \alpha_k\sigma _k^2}}{{{p_s}}}z}}}}{{1 + z}}\mathrm{d}} z\), and \(C_1^\prime\triangleq\int_0^\infty  {\frac{{{e^{ - \frac{{\left( {{\lambda _k} + {\lambda _{\bar k}}} \right)\alpha_k\sigma _k^2}}{{{p_w}}}z}}}}{{1 + z}}\mathrm{d}} z\), which are obtained by substituting \eqref{eq:conditional CDF of Gamma_k in OMA-II} and \eqref{eq:conditional CDF of tilde Gamma_k in OMA-II} for \({F_{{\Gamma _{k|X \ge Y}}}}({{z}})\) and \({F_{{\tilde\Gamma _{k|X<Y}}}}({{z}})\), respectively. Then by directly applying \cite[Eq.~(3.352.4)]{gradshteyn}, \(A_1^\prime=-f(\frac{{\left( {{\lambda _k} + {\lambda _{\bar k}}} \right)\alpha_k\sigma _k^2}}{{{p_s}}})\), \(B_1^\prime=-f(\frac{{{{\lambda _k} }\alpha_k\sigma _k^2}}{{{p_s}}})\) and \(C_1^\prime=-f(\frac{{\left( {{\lambda _k} + {\lambda _{\bar k}}} \right)\alpha_k\sigma _k^2}}{{{p_w}}})\) are derived, which completes the proof for Proposition~\ref{prop:ER for the kth user in OMA-II}.

\section{}\label{appendix:proof of optimal power allocation for DLT OMA-II}
Following similar analysis as for Proposition~\ref{prop:optimal power allocation for DLT NOMA}, the minimum of \(\bar{\mathcal{L}}_2^{\rm OMA\text{-}II}(p_k,p_{\bar k},\alpha_k)\) is obtained by comparing all possible combinations of outage occurrences for \(\mathcal{U}_{k}\) and \(\mathcal{U}_{\bar k}\) (c.f.~Table~\ref{table:combinations of U1 and U2's OMA-II decoding strategies}). In \eqref{eq:case-specified optimal power allocation for DLT OMA-II under full CSIT}, \(p_{i,k}\) and \(p_{i,\bar k}\), \(i=1,2,3,4\), are respectively the minimum  power required to have both of the users suspend their transmission, only \(\mathcal{U}_{k}\) or \(\mathcal{U}_{\bar k}\) supported, and both of the users simultaneously served.

\section{}\label{appendix:proof of power saving for DLT NOMA over OMA-II}
Note from Lemma~\ref{lemma:power saving for DLT NOMA over OMA-II} that to prove \(P_{\rm O2}^\ast\ge P_{\rm N}^\ast\), it is sufficient to show that \(\sum_{i=1}^K\frac{(2^{\frac{\bar R_i}{\alpha_i}}-1)\alpha_i}{g_i}\ge P_{\rm N}^\ast\) holds for any \(\alpha_i\)'s such that \(\sum_{i=1}^K\alpha_i=1\). By variable transformation of \(i\leftarrow(K-i)\), it follows that 
\(P_{\rm N}^\ast=\sum_{i=1}^K\frac{(2^{\bar R_i}-1)2^{\sum_{l=i+1}^K\bar R_l}}{g_i}\). 

First, denoting \((2^{\frac{\bar R_i}{\alpha_i}}-1)\alpha_i\) by \(a_i\), and \((2^{\bar R_i}-1)2^{\sum_{l=i+1}^K\bar R_l}
\) by \(b_i\), \(\forall i\), we prove that \(\sum_{i=j}^Ka_i\ge \sum_{i=j}^Kb_i\) holds for \(\forall j=1, \ldots,K\). Expand \(\sum_{i=j}^Kb_i\) as follows:
\begin{align} 
\left\{\begin{array} {cl}
2^{\sum_{l=j}^K\bar R_l}-2^{\sum_{l=j+1}^K\bar R_l}, & i=j\\
2^{\sum_{l=j+1}^K\bar R_l}-2^{\sum_{l=j+2}^K\bar R_l}, & i=j+1\\
\vdots & i=j+2,\ldots,K-1\\
2^{\bar R_K}-1. & i=K
\end{array}\right. \label{eq:expansion of sum of i=j:Kb_i}
\end{align}
By summing-up the LHS of \eqref{eq:expansion of sum of i=j:Kb_i}, \(\sum_{i=j}^Kb_i\) is simplified as \(2^{\sum_{i=j}^K\bar R_i}-1\). Defining a function \(f_0(x)=2^x-1\), let \(x_i=0\) for \(i=1,\ldots,j-1\), and \(x_i=\frac{\bar R_i}{\alpha_i}\) for \(i=j,\ldots,K\). Then by applying Jensen's inequality due to the convexity of \(f_0(x)\), it follows that
\begin{multline}
\sum_{i=j}^Ka_i = \sum_{i=1}^K\alpha_if_0(x_i)\\
\ge f_0\left(\sum_{i=1}^K\alpha_ix_i \right )=f_0\left(\sum_{i=j}^K\alpha_ix_i \right )=\sum_{i=j}^Kb_i. \label{eq:Jensen's inequality} 
\end{multline}

Next, define \(\frac{1}{g_i}\) by \(c_i\), \(\forall i\), it is easily verified that \(0<c_1\le c_2\le\ldots\le c_K\). By applying \cite[{\bf Lemma 4}]{chen17mathematical}, we conclude that \(\sum_{i=1}^K\frac{(2^{\frac{\bar R_i}{\alpha_i}}-1)\alpha_i}{g_i}\ge\sum_{i=1}^K\frac{(2^{\bar R_i}-1)2^{\sum_{l=i+1}^K\bar R_l}}{g_i}
=P_{\rm N}^\ast\), which completes the proof for Lemma~\ref{lemma:power saving for DLT NOMA over OMA-II}.

\section{}\label{appendix:proof of case-specific sigma_k prime NOMA under partial CSIT}
The piece-wise presentation of \(\mb{E}_\nu[X_k^{\prime\rm NOMA}(\nu)]\) is caused by the range of the parameters. To illustrate as an example, take the second term of \eqref{eq:sigma_k prime NOMA under partial CSIT} as an example and express it as follows:
\begin{align}
&\Pr\bigg\{\log_2\left(1+\frac{p_sX}{\sigma_k^2} \right)<\bar R_k,\log_2\left(1+\frac{p_wX}{p_sX+\sigma_k^2} \right )\ge\bar R_{\bar k},\notag\\
&\ ~~~~ X>Y \bigg\}\notag\\
&=\Pr\left\{X<\varepsilon_{k,1},\left(p_w-p_s\tau_{\bar k}\right )X\ge\sigma_k^2\tau_{\bar k},X>Y\right\}\notag\\
&=\left\{\kern-4pt\begin{array}{l} \Pr\left\{X<\varepsilon_{k,1},X\ge\varepsilon_{k,4},X>Y\right\}, \ {\rm if}\ \frac{p_s}{p_w}<\frac{1}{\tau_{\bar k}},\\
0,\ {\rm otherwise.}
\end{array}\right. \label{eq:outage probability after successful SIC for user k}
\end{align} Further, the first case in \eqref{eq:outage probability after successful SIC for user k} implies the following two sub-cases:\\
{\bf Case 1}: \(\frac{p_s}{p_w}<\min\left\{\frac{\tau_k}{\tau_{\bar k}(\tau_k+1)},\frac{1}{\tau_{\bar k}}\right\}\)
\begin{multline}
\Pr\left\{X<\varepsilon_{k,1},X\ge\varepsilon_{k,4},X>Y\right\}=\\
\Pr\left\{\varepsilon_{k,4}\le X<\varepsilon_{k,1},X>Y\right\}; \label{eq:the first subcase}
\end{multline} 
{\bf Case 2}: \(\frac{p_s}{p_w}\ge\min\left\{\frac{\tau_k}{\tau_{\bar k}(\tau_k+1)},\frac{1}{\tau_{\bar k}}\right\}\)
\begin{align}
\Pr\left\{X<\varepsilon_{k,1},X\ge\varepsilon_{k,4},X>Y\right\}=0. \label{eq:the second subcase}
\end{align}
Hence, in the case of \(\frac{p_s}{p_w}<\min\{\Myfrac{\tau_k}{\tau_{\bar k}(\tau_k+1)},\Myfrac{1}{\tau_{\bar k}}\}\), after some manipulations, the second term of \eqref{eq:sigma_k prime NOMA under partial CSIT} turns out to be \(e^{-\lambda_k\varepsilon_{k,4}}-e^{-\lambda_k\varepsilon_{k,1}}+\frac{\lambda_k}{\lambda_k+\lambda_{\bar k}}(e^{-(\lambda_k+\lambda_{\bar k})\varepsilon_{k,1}}-e^{-(\lambda_k+\lambda_{\bar k})\varepsilon_{k,4}})\) (c.f.~\eqref{eq:the first subcase}), and otherwise zero (c.f.~\eqref{eq:the second subcase}). By analogy, the first and the third term of \eqref{eq:sigma_k prime NOMA under partial CSIT} can also be analysed piece-wisely. Finally, we arrive at \eqref{eq:case-specific zeta_k prime NOMA under partial CSIT} combining all possible cases.
\end{appendices}

%\linespread{1.3}
%the map from commands to actual font sizes can be found at https://en.wikibooks.org/wiki/LaTeX/Fonts
%\balance
\bibliographystyle{IEEEtran}
\IEEEtriggeratref{15} % This fires a new column at the given BibTeX reference number
\bibliography{NOMA_ref}

% Generated by IEEEtran.bst, version: 1.14 (2015/08/26)
\begin{thebibliography}{10}
\providecommand{\url}[1]{#1}
\csname url@samestyle\endcsname
\providecommand{\newblock}{\relax}
\providecommand{\bibinfo}[2]{#2}
\providecommand{\BIBentrySTDinterwordspacing}{\spaceskip=0pt\relax}
\providecommand{\BIBentryALTinterwordstretchfactor}{4}
\providecommand{\BIBentryALTinterwordspacing}{\spaceskip=\fontdimen2\font plus
\BIBentryALTinterwordstretchfactor\fontdimen3\font minus
  \fontdimen4\font\relax}
\providecommand{\BIBforeignlanguage}[2]{{%
\expandafter\ifx\csname l@#1\endcsname\relax
\typeout{** WARNING: IEEEtran.bst: No hyphenation pattern has been}%
\typeout{** loaded for the language `#1'. Using the pattern for}%
\typeout{** the default language instead.}%
\else
\language=\csname l@#1\endcsname
\fi
#2}}
\providecommand{\BIBdecl}{\relax}
\BIBdecl

\bibitem{xing2017WCNC}
H.~Xing, Y.~Liu, A.~Nallanathan, and Z.~Ding, ``Sum-rate maximization
  guaranteeing user fairness for {NOMA} in fading channels,'' in \emph{Proc.
  IEEE Wireless Communications and Networking Conference (WCNC)}, Barcelona,
  Spain, Apr. 2018.

\bibitem{Saito13VTC}
Y.~Saito, Y.~Kishiyama, A.~Benjebbour, T.~Nakamura, A.~Li, and K.~Higuchi,
  ``Non-orthogonal multiple access ({NOMA}) for cellular future radio access,''
  in \emph{Proc. IEEE Vehicular Technology Conference (VTC Spring)}, Dresden,
  Germany, June 2013.

\bibitem{Fei16MaMIMO}
H.~Xie, B.~Wang, F.~Gao, and S.~Jin, ``A full-space spectrum-sharing strategy
  for massive {MIMO} cognitive radio systems,'' \emph{{IEEE} J. Sel. Areas
  Commun.}, vol.~34, no.~10, pp. 2537--2549, Oct. 2016.

\bibitem{Dai15ComMag}
L.~Dai, B.~Wang, Y.~Yuan, S.~Han, C.~l.~I, and Z.~Wang, ``Non-orthogonal
  multiple access for {5G}: solutions, challenges, opportunities, and future
  research trends,'' \emph{{IEEE} Commun. Mag.}, vol.~53, no.~9, pp. 74--81,
  Sept. 2015.

\bibitem{Zhiguo2017Mag}
Z.~Ding, Y.~Liu, J.~Choi, Q.~Sun, M.~Elkashlan, C.-L. I, and H.~V. Poor,
  ``Application of non-orthogonal multiple access in {LTE} and {5G} networks,''
  \emph{{IEEE} Commun. Mag.}, vol.~55, no.~2, pp. 185--191, Feb. 2017.

\bibitem{LTE2015}
3rd Generation Partnership Project~({3GPP}), ``Study on downlink multiuser
  superposition transmission for {LTE},'' Mar. 2015.

\bibitem{Zhang16TV}
L.~Zhang, W.~Li, Y.~Wu, X.~Wang, S.~I. Park, H.~M. Kim, J.~Y. Lee, P.~Angueira,
  and J.~Montalban, ``Layered-division-multiplexing: theory and practice,''
  \emph{{IEEE} Trans. on Broadcast.}, vol.~62, no.~1, pp. 216--232, Mar. 2016.

\bibitem{ding2014performance}
Z.~Ding, Z.~Yang, P.~Fan, and H.~V. Poor, ``On the performance of
  non-orthogonal multiple access in {5G} systems with randomly deployed
  users,'' \emph{{IEEE} Signal Process. Lett.}, vol.~21, no.~12, pp.
  1501--1505, 2014.

\bibitem{Jinho2014Comp}
J.~Choi, ``Non-orthogonal multiple access in downlink coordinated two-point
  systems,'' \emph{{IEEE} Commun. Lett.}, vol.~18, no.~2, pp. 313--316, Feb.
  2014.

\bibitem{Shin2017NOMA}
W.~Shin, M.~Vaezi, B.~Lee, D.~J. Love, J.~Lee, and H.~V. Poor, ``Coordinated
  beamforming for multi-cell {MIMO-NOMA},'' \emph{{IEEE} Commun. Lett.},
  vol.~21, no.~1, pp. 84--87, Jan. 2017.

\bibitem{ding2015cooperative}
Z.~Ding, M.~Peng, and H.~V. Poor, ``Cooperative non-orthogonal multiple access
  in {5G} systems,'' \emph{{IEEE} Commun. Lett.}, vol.~19, no.~8, pp.
  1462--1465, Aug. 2015.

\bibitem{yuanwei_JSAC_2015}
Y.~Liu, Z.~Ding, M.~Elkashlan, and H.~V. Poor, ``Cooperative non-orthogonal
  multiple access with simultaneous wireless information and power transfer,''
  \emph{{IEEE} J. Sel. Areas Commun.}, vol.~34, no.~4, pp. 938--953, Apr. 2016.

\bibitem{Timotheou:2015}
S.~Timotheou and I.~Krikidis, ``Fairness for non-orthogonal multiple access in
  {5G} systems,'' \emph{{IEEE} Signal Process. Lett.}, vol.~22, no.~10, pp.
  1647--1651, Oct. 2015.

\bibitem{Yuanwei2016NOMA}
Y.~Liu, M.~Elkashlan, Z.~Ding, and G.~K. Karagiannidis, ``Fairness of user
  clustering in {MIMO} non-orthogonal multiple access systems,'' \emph{{IEEE}
  Commun. Lett.}, vol.~20, no.~7, pp. 1465--1468, Jul. 2016.

\bibitem{chen17mathematical}
Z.~Chen, Z.~Ding, X.~Dai, and R.~Zhang, ``An optimization perspective of the
  superiority of {NOMA} compared to conventional {OMA},'' \emph{{IEEE} Trans.
  Signal Process.}, vol.~65, no.~19, pp. 5191--5202, Oct 2017.

\bibitem{Di2016TWC}
B.~Di, L.~Song, and Y.~Li, ``Sub-channel assignment, power allocation, and user
  scheduling for non-orthogonal multiple access networks,'' \emph{{IEEE} Trans.
  Wireless Commun.}, vol.~15, no.~11, pp. 7686--7698, Nov. 2016.

\bibitem{Sun17FD-NOMA}
Y.~Sun, D.~W.~K. Ng, Z.~Ding, and R.~Schober, ``Optimal joint power and
  subcarrier allocation for full-duplex multicarrier non-orthogonal multiple
  access systems,'' \emph{{IEEE} Trans. Commun.}, vol.~65, no.~3, pp.
  1077--1091, Mar. 2017.

\bibitem{Lifang01EC}
L.~Li and A.~Goldsmith, ``Capacity and optimal resource allocation for fading
  broadcast channels .{I}. ergodic capacity,'' \emph{{IEEE} Trans. Inf.
  Theory}, vol.~47, no.~3, pp. 1083--1102, Mar. 2001.

\bibitem{Lifang01OC}
------, ``Capacity and optimal resource allocation for fading broadcast
  channels .{II}. outage capacity,'' \emph{{IEEE} Trans. Inf. Theory}, vol.~47,
  no.~3, pp. 1103--1127, Mar. 2001.

\bibitem{xing2016secrecySWIPT}
H.~Xing, L.~Liu, and R.~Zhang, ``Secrecy wireless information and power
  transfer in fading wiretap channel,'' \emph{{IEEE} Trans. Veh. Technol.},
  vol.~65, no.~1, pp. 180--190, Jan. 2016.

\bibitem{Nihar03minimum}
N.~Jindal and A.~Goldsmith, ``Capacity and optimal power allocation for fading
  broadcast channels with minimum rates,'' \emph{{IEEE} Trans. Inf. Theory},
  vol.~49, no.~11, pp. 2895--2909, Nov. 2003.

\bibitem{Dirk1975thesis}
D.~Hughes-Hartogs, ``The capacity of a degraded spectral gaussian broadcast
  channel,'' Ph.D. dissertation, Inform. Syst. Lab., Ctr. Syst. Res., Stanford
  Univ., Stanford, CA, Jul. 1975.

\bibitem{MUST2015}
\emph{Study on Downlink Multiuser Superposition Transmission ({MUST}) for {LTE}
  ({Release 13})}, 3GPP document TR 36.859, Dec. 2015.

\bibitem{yongpeng15finite-alphabet}
Y.~Wu, C.~K. Wen, C.~Xiao, X.~Gao, and R.~Schober, ``Linear precoding for the
  {MIMO} multiple access channel with finite alphabet inputs and statistical
  {CSI},'' \emph{{IEEE} Trans. Wireless Commun.}, vol.~14, no.~2, pp. 983--997,
  Feb. 2015.

\bibitem{Zheng17JSAC}
Z.~Dong, H.~Chen, J.~K. Zhang, and L.~Huang, ``On non-orthogonal multiple
  access with finite-alphabet inputs in {Z}-channels,'' \emph{{IEEE} J. Sel.
  Areas Commun.}, vol.~35, no.~12, pp. 2829--2845, Dec. 2017.

\bibitem{Stephen98delay-limited}
S.~V. Hanly and D.~N.~C. Tse, ``Multiaccess fading channels. ii. delay-limited
  capacities,'' \emph{{IEEE} Trans. Inf. Theory}, vol.~44, no.~7, pp.
  2816--2831, Nov 1998.

\bibitem{Hina17ModelingPCP}
H.~Tabassum, E.~Hossain, and J.~Hossain, ``Modeling and analysis of uplink
  non-orthogonal multiple access in large-scale cellular networks using
  {Poisson} cluster processes,'' \emph{{IEEE} Trans. Commun.}, vol.~65, no.~8,
  pp. 3555--3570, Aug. 2017.

\bibitem{Yu2006}
W.~Yu and R.~Lui, ``Dual methods for nonconvex spectrum optimization of
  multicarrier systems,'' \emph{{IEEE} Trans. Commun.}, vol.~54, no.~7, pp.
  1310--1322, July 2006.

\bibitem{rockafellar1997convex}
R.~T. Rockafellar, \emph{Convex Analysis}.\hskip 1em plus 0.5em minus
  0.4em\relax Princeton Univ. Press, 1997.

\bibitem{EE364b}
\BIBentryALTinterwordspacing
S.~Boyd, ``Lecture notes for {EE}364b: {Convex Optimization II}.'' [Online].
  Available: \url{https://stanford.edu/class/ee364b/lectures.html}
\BIBentrySTDinterwordspacing

\bibitem{Liu2016TVT}
Y.~Liu, Z.~Ding, M.~Elkashlan, and J.~Yuan, ``Non-orthogonal multiple access in
  large-scale underlay cognitive radio networks,'' \emph{{IEEE} Trans. Veh.
  Technol.}, vol.~65, no.~12, pp. 10\,152--10\,157, Dec. 2016.

\bibitem{abramowitz1972handbook}
M.~Abramowitz and I.~A. Stegun, \emph{Handbook of Mathematical Functions: with
  Formulas, Graphs, and Mathematical Tables}, 9th~ed.\hskip 1em plus 0.5em
  minus 0.4em\relax Mineola, NY, USA: Dover Publication, Inc., 1972.

\bibitem{gradshteyn}
I.~S. Gradshteyn and I.~M. Ryzhik, \emph{Table of Integrals, Series and
  Products}, 6th~ed.\hskip 1em plus 0.5em minus 0.4em\relax New York, NY, USA:
  Academic Press, 2000.

\end{thebibliography}

\end{document}